\documentclass[11pt,a4paper]{article}

\usepackage[margin=1in]{geometry}
\usepackage{libertinus}          
\usepackage{microtype}
\usepackage{parskip}
\usepackage{titlesec}
\titleformat{\section}
  {\normalfont\Large\bfseries}
  {\thesection}{0.75em}{}[\vspace{0.25ex}\titlerule]
\titleformat{\subsection}
  {\normalfont\large\bfseries}
  {\thesubsection}{0.6em}{}
\titleformat{\subsubsection}
  {\normalfont\normalsize\bfseries}
  {\thesubsubsection}{0.55em}{}
\titlespacing*{\section}{0pt}{2.2ex plus .8ex minus .2ex}{1.0ex}
\titlespacing*{\subsection}{0pt}{1.8ex plus .7ex minus .2ex}{0.8ex}
\titlespacing*{\subsubsection}{0pt}{1.4ex plus .5ex minus .2ex}{0.6ex}

\usepackage{enumitem}
\setlist{nosep,leftmargin=*}

\usepackage[hidelinks]{hyperref}
\usepackage{orcidlink}

\usepackage{comment} 
\usepackage{caption}
\usepackage{amsmath}
\usepackage{float}
\usepackage[utf8x]{inputenc}
\usepackage{amsmath,amssymb}
\usepackage{xcolor}
\usepackage{mathtools}
\usepackage{tabulary}
\usepackage{amsthm}
\usepackage{latexsym}
\usepackage[mathcal]{eucal}
\usepackage{amscd}
\usepackage{graphicx}
\usepackage{amsfonts}
\usepackage{amscd}
\usepackage{MnSymbol}
\usepackage{setspace}

\usepackage[english]{babel}
\usepackage{listings}
\usepackage{calligra}
\usepackage{titlesec}
\usepackage{booktabs}

\theoremstyle{remark}

\usepackage{titling} 

\pretitle{%
  \begin{center}%
  \rule{\textwidth}{1.4pt}\vspace{1.7ex}\\
  \LARGE\bfseries
}
\posttitle{%
  \rule{\textwidth}{1.4pt}
  \end{center}\vspace{1ex}
}

\title{\textbf{\Large  The Universe Learning Itself: On the Evolution of Dynamics from the\\ Big Bang to Machine Intelligence
}}

\author{
    \large 
    Pradeep Singh\orcidlink{0000-0002-5372-3355}\thanks{Email: \texttt{pradeep.cs@sric.iitr.ac.in}},  Mudasani Rushikesh\orcidlink{0009-0008-1439-780X}\thanks{Email: \texttt{mudasani\_r@cs.iitr.ac.in}},  Bezawada Sri Sai Anurag\orcidlink{0009-0007-7879-8787}\thanks{Email: \texttt{bezawada\_ssa@cs.iitr.ac.in}},\\    
    Balasubramanian Raman\orcidlink{0000-0001-6277-6267}\thanks{Email: \texttt{bala@cs.iitr.ac.in}}\vspace{0.2cm}\\
    \begin{minipage}[t]{0.5\textwidth}
    \centering
    \small Department of Computer Science and Engineering\\
    \small Indian Institute of Technology Roorkee\\
    \small Roorkee-247667, India
    \end{minipage}
}

\date{}

\newenvironment{keywords}{%
  \par\vspace{1.5ex}%
  \noindent\textbf{Keywords: }%
}{%
  \par\vspace{0.6ex}%
}

\begin{document}\maketitle

\begin{abstract}
We develop a unified, dynamical-systems narrative of the universe that traces a continuous chain of structure formation from the Big Bang to contemporary human societies and their artificial learning systems. Rather than treating cosmology, astrophysics, geophysics, biology, cognition, and machine intelligence as disjoint domains, we view each as successive regimes of dynamics on ever-richer state spaces, stitched together by phase transitions, symmetry-breaking events, and emergent attractors. Starting from inflationary field dynamics and the growth of primordial perturbations, we describe how gravitational instability sculpts the cosmic web, how dissipative collapse in baryonic matter yields stars and planets, and how planetary-scale geochemical cycles define long-lived nonequilibrium attractors. Within these attractors, we frame the origin of life as the emergence of self-maintaining reaction networks, evolutionary biology as flow on high-dimensional genotype–phenotype–environment manifolds, and brains as adaptive dynamical systems operating near critical surfaces. Human culture and technology—including modern machine learning and artificial intelligence—are then interpreted as symbolic and institutional dynamics that implement and refine \emph{engineered} learning flows which recursively reshape their own phase space. Throughout, we emphasize recurring mathematical motifs—instability, bifurcation, multiscale coupling, and constrained flows on measure-zero subsets of the accessible state space. Our aim is  not to present  any new cosmological or biological model, but a cross-scale, theoretical perspective: a way of reading the universe’s history as the evolution of dynamics itself, culminating (so far) in biological and artificial systems capable of modeling, predicting, and deliberately perturbing their own future trajectories.

\end{abstract}


\section{Introduction}

Contemporary science offers an astonishing empirical narrative: from the hot, dense plasma of the early universe to galaxies and stars, from chemically active planets to cells, organisms, brains, and finally to cultures, technologies, and artificial learning systems that model the cosmos which produced them \cite{Chaisson2001,Chaisson2011,JordanMitchell2015,LeCun2015}. Yet this story is usually told as a sequence of disciplinary episodes---cosmology, astrophysics, geophysics, biology, neuroscience, cognitive science, machine learning---each with its own language and models. In this work we ask a different question: \emph{Can we read the entire history of the universe as a single, cross-scale dynamical trajectory, structured by a small set of recurring mathematical motifs?}

Our starting point is the view, central to dynamical systems theory, that the fundamental object of interest is a state space, a flow on that state space, and the invariant sets and bifurcations that organize long-term behaviour. Over the last decades, this viewpoint has reshaped our understanding of nonequilibrium structure formation in physics \cite{NicolisPrigogine1977}, self-organization and critical phenomena in complex systems \cite{Bak1996}, the emergence of autocatalytic chemistry and early life \cite{Kauffman1993,Hordijk2019,Steel2019}, and the collective dynamics of neural tissue poised near criticality \cite{BeggsPlenz2003,Timme2016}. In parallel, it has become a dominant language for machine learning and artificial intelligence: training deep networks is understood as gradient flow on high-dimensional loss landscapes, while inference and control are seen as dynamical processes on representation manifolds and policy spaces \cite{MezardMontanari2009,Saxe2014,AdvaniSaxe2017,Schoenholz2017,Schmidhuber2015}. At the same time, “cosmic evolution” and “big history” programmes have emphasized the continuity of structure formation across cosmological, astrophysical, biological, and cultural scales, often using energetic metrics such as energy rate density as a unifying quantity \cite{Chaisson2011}. What is still missing, we argue, is a \emph{systematic dynamical-systems reading} of this full cross-scale history that also incorporates artificial learning systems as a late-stage continuation of the same motifs.

In such a reading, the early universe is not merely “initial conditions”, but a regime of field dynamics whose fluctuations and instabilities define the measure on later trajectories. Gravitational amplification of these fluctuations becomes a concrete example of linear instability giving way to nonlinear pattern formation: small Gaussian perturbations grow, couple, and collapse into the cosmic web of halos, filaments, and voids, selecting a highly structured subset of the universe’s naively accessible microstates. Gas cooling and angular-momentum transport within halos can then be seen as dissipative flows driving matter toward stellar and planetary quasi-attractors; stellar evolution traces predictable paths through a constrained manifold of nuclear reaction networks. From this perspective, “structure” is always shorthand for the geometry of attractors, invariant manifolds, and basins in appropriate state spaces.

On planetary scales, the same language captures climate, geochemical cycles, and tectonics as coupled, multiscale dynamical systems with feedbacks and multiple attractors \cite{NicolisPrigogine1977}. Habitable conditions correspond not just to particular parameter values, but to regions of parameter space where long-lived, far-from-equilibrium attractors exist and are structurally stable to perturbations. Within such regimes, prebiotic chemistry explores enormous reaction networks; the origin of life can be framed as a phase transition in which an autocatalytic, self-maintaining subnetwork becomes dynamically autonomous, carving out a bounded region of state space that resists diffusion and decay \cite{Kauffman1993,Hordijk2019,Steel2019}. Biological evolution then appears as flow on coupled genotype–phenotype–environment manifolds, with adaptive peaks, ridges, and neutral networks providing the geometric scaffolding for evolutionary trajectories \cite{Nowak2006}.

Nervous systems add a further layer: adaptive controllers whose own state spaces are tuned so that dynamics lie near critical surfaces, balancing stability and flexibility. Empirical evidence for neural criticality suggests that cortical networks operate near phase transitions, where dynamic range, information transmission, and complexity are optimized \cite{BeggsPlenz2003,Timme2016}. Brains thus instantiate dynamical systems that model, predict, and control other dynamical systems, including themselves. With the emergence of language, culture, and technology, a new class of “symbolically extended” dynamics appears: networks of agents, institutions, and artefacts whose interactions recursively modify their own effective state spaces and control parameters. Within this symbolic layer, modern machine learning and AI define explicitly engineered learning dynamics: gradient-based optimization in parameter spaces, reinforcement learning in environment–policy loops, and large-scale foundation models that, in a literal sense, participate in modelling and shaping the very socio-technical systems in which they are embedded \cite{JordanMitchell2015,LeCun2015,Schmidhuber2015,SuttonBarto2018}. From this angle, the recent explosion of artificial intelligence and global communication infrastructures can be interpreted as yet another bifurcation in the universe’s self-organized dynamics.

The aim of this paper is not to propose a new quantitative model of any specific subsystem—cosmological, biological, cognitive, or artificial—nor to claim predictive power across all scales. Rather, our goal is \emph{conceptual}: to articulate a coherent dynamical-systems narrative that places standard episodes of cosmic and biological history, together with the emergence of engineered learning systems, into a single, unified mathematical frame. We identify a small set of recurring motifs—instability and amplification, symmetry breaking, emergence of new slow variables, formation of attractor manifolds of increasing effective dimensionality, and the appearance of systems that adaptively navigate their own state spaces—and track how these motifs reappear from the Big Bang to contemporary human and machine intelligence.

More concretely, we proceed as follows. In Section~\ref{sec:inflation} we discuss the early universe and inflationary dynamics as the birth of the effective phase space and the distribution of initial perturbations. Section~\ref{sec:cosmic_web} recasts gravitational structure formation as flow from linear modes to the nonlinear cosmic web. Section~\ref{sec:stellar_planetary} frames stars and planetary systems as dissipative attractors in coupled radiative–hydrodynamic dynamics. Section~\ref{sec:planetary_dynamics} treats Earth-like planets as multiscale dynamical systems with climate and geochemical feedbacks. Section~\ref{sec:origin_life} and Section~\ref{sec:evolution} reinterpret the origin of life and biological evolution in terms of autocatalytic sets and flows on genotype–phenotype–environment manifolds. Section~\ref{sec:brains} and Section~\ref{sec:culture} move to neural, cognitive, and cultural dynamics, with an emphasis on criticality, self-referential system design, and the emergence of machine learning and AI as explicitly designed learning dynamics. Finally, Section~\ref{sec:synthesis} synthesizes these threads, discussing arrows of time, the rarity or typicality of self-modeling systems, and open questions about whether the emergence of coupled biological and artificial intelligences is a generic late-time attractor or a finely tuned contingency.

By treating “from Big Bang to brain and beyond” as a single, evolving tapestry of dynamics, we hope to make two contributions. First, we provide a precise mathematical vocabulary for talking about cross-scale structure formation that complements existing energetic and informational metrics \cite{Chaisson2011,Georgiev2015}. Second, we aim to provoke further work at the interfaces between cosmology, complex systems, the sciences of life and mind, and machine learning: if the universe’s history can indeed be read as the evolution of its own effective dynamics, then the appearance of systems capable of understanding, extending, and steering those dynamics—biological and artificial—may not be an epilogue, but a central chapter in the story.

\section{Inflation, Initial Conditions, and the Birth of Phase Space}
\label{sec:inflation}

In standard cosmology texts, cosmic inflation is introduced as a clever patch for a handful of puzzles in the hot Big Bang model: the near-uniformity of the cosmic microwave background (CMB) across causally disconnected regions, the near-flatness of spatial geometry, and the absence of relics such as magnetic monopoles \cite{Guth1981,Mukhanov2005}. In this section we adopt a different emphasis. We treat the inflationary era not only as a solution to specific fine-tuning problems, but as the \emph{birth of the effective phase space} in which the rest of cosmic history unfolds. Inflation dynamically selects a low-dimensional manifold of states (the slow-roll attractor), specifies a stochastic measure on perturbations around that manifold, and thus determines which regions of the universe’s vastly larger kinematic state space are actually explored.

\subsection{Inflation as a dynamical attractor}

The simplest models of inflation posit a scalar degree of freedom---the inflaton---with canonical kinetic term and a potential $V(\phi)$, minimally coupled to gravity. In a spatially flat Friedmann--Lemaître--Robertson--Walker (FLRW) background, the coupled gravity--field equations reduce to a low-dimensional dynamical system for the homogeneous mode $\phi(t)$ and Hubble parameter $H(t)$:
\begin{equation}
\dot{\phi} = \pi, \qquad
\dot{\pi} = - 3H\pi - V'(\phi), \qquad
H^2 = \frac{8\pi G}{3}\Bigl(\tfrac{1}{2}\pi^2 + V(\phi)\Bigr),
\end{equation}
where $\pi$ denotes the field momentum and overdots denote derivatives with respect to cosmic time. Slow-roll inflation corresponds to trajectories in the $(\phi,\pi)$ phase plane for which the friction term $3H\pi$ dominates the acceleration $\dot{\pi}$ and the potential energy dominates the kinetic energy, so that $\pi^2 \ll V(\phi)$ and $\dot{\phi}\approx -V'(\phi)/3H$ \cite{Linde1983,Mukhanov2005}. 

From a dynamical-systems point of view, this regime defines an \emph{attractor}: a one-dimensional slow manifold in the $(\phi,\pi)$ phase space to which a broad range of initial conditions are drawn. Anisotropies, spatial curvature, and many details of the pre-inflationary state are exponentially diluted; trajectories that start with wildly different values of $\pi$ or small amounts of shear rapidly converge onto the same slow-roll track as long as certain coarse conditions are met (e.g., the potential is sufficiently flat over a region in field space). In other words, inflation is a mechanism by which the universe dynamically forgets most of its initial microstate, collapsing a high-dimensional space of possibilities onto a much smaller effective state space. This is a concrete instance of a general motif we will see repeatedly: fast, dissipative directions contracting volumes in phase space and leaving only a few slow variables to organize long-term behaviour.

This attractor nature also reframes the ``fine-tuning'' narrative. Rather than requiring exquisitely tuned initial conditions to obtain a homogeneous, flat, and isotropic universe, inflation renders such conditions typical within the basin of attraction of slow-roll dynamics. In the language of dynamical systems, the physically relevant measure on cosmic initial conditions is concentrated on the attractor manifold and its small perturbations, not spread uniformly over a naive kinematic state space. The question of why our universe lies near this manifold becomes, at least partly, the question of why we inhabit a region of space-time that underwent sufficient inflation; the statistical weight of such regions is a dynamical output, not an arbitrary prior assumption \cite{Linde1986,Linde1984}.

\subsection{Quantum fluctuations and the seeding of structure}

While the homogeneous slow-roll attractor sets the background trajectory, it is the \emph{fluctuations} around this trajectory that encode the seeds of all later structure. Quantum field theory in curved space-time predicts that the inflaton and metric are subject to vacuum fluctuations; during inflation, these fluctuations are stretched to super-Hubble scales and ``freeze out'' as classical perturbations in the curvature and energy density \cite{Mukhanov2005}. The result is a random field---often represented by the comoving curvature perturbation~$\zeta(\mathbf{x})$---whose statistics are determined by the inflationary dynamics.

For single-field slow-roll models, these perturbations form an almost Gaussian random field with an almost scale-invariant power spectrum,
\begin{equation}
\mathcal{P}_{\zeta}(k) \approx A_s \left(\frac{k}{k_\ast}\right)^{n_s - 1},
\end{equation}
where $A_s$ is the amplitude, $n_s$ the spectral index, and $k_\ast$ a pivot scale. Observations of the CMB anisotropies by \emph{Planck} and other experiments tightly constrain these numbers; for example, the latest \emph{Planck} analysis gives $n_s \approx 0.965$ with no strong evidence for running or large non-Gaussianity, consistent with a broad class of slow-roll inflationary models \cite{Planck2018Inflation,Planck2018Params}. 

From our perspective, these results are more than just parameter estimates. They tell us that inflation generates a very specific \emph{probability distribution} over initial perturbation states: effectively, a measure on the space of possible cosmological trajectories emerging from the end of inflation. Each Fourier mode $k$ of the curvature perturbation is drawn from a nearly independent Gaussian with variance set by $\mathcal{P}_{\zeta}(k)$. The realization our universe happens to sample is encoded in the CMB temperature and polarization anisotropies and in the large-scale distribution of galaxies. Thus, inflation does not merely smooth the universe; it populates the effective phase space with a well-characterized ensemble of small inhomogeneities that will later be amplified by gravitational instability.

A subtle, but important, dynamical point is the quantum-to-classical transition. The same exponential stretching that moves modes outside the Hubble radius also decoheres them: different field configurations become effectively classical, with interference suppressed when traced over unobserved degrees of freedom. Coarse-graining over sub-Hubble fluctuations yields an emergent classical stochastic process for the long-wavelength modes, often captured by Langevin- or Fokker--Planck-type equations in ``stochastic inflation'' approaches \cite{Starobinsky1986}. In this way, inflation provides a concrete example of how classical stochastic dynamics can emerge from underlying quantum dynamics via expansion and coarse-graining.

\subsection{Initial conditions as a dynamically generated measure}

If we view the rest of cosmic history as a trajectory in a vast state space of density and velocity fields, then the output of inflation is a \emph{probability measure} on this state space at some early time: a Gaussian random field characterized by a small set of parameters $(A_s, n_s, \ldots)$, plus a nearly homogeneous and isotropic background. The usual language of ``initial conditions'' thus disguises a deeper dynamical story. Rather than positing a particular microstate by fiat, we posit a relatively simple dynamical law for the early universe (slow-roll inflation), and we derive from it a restricted ensemble of initial microstates. 

In dynamical-systems terms, one can think of inflation as performing two operations on the primordial phase space. First, the attractor dynamics collapses an enormous set of possible pre-inflationary configurations onto a lower-dimensional slow manifold parameterized by a few background variables (e.g., the value of $\phi$ when observable modes exit the Hubble radius). Second, quantum fluctuations around this manifold are linearly amplified and stretched into classical perturbations, generating a measure on small deviations from the attractor. These deviations are small in amplitude but rich in structure; they are precisely the degrees of freedom that will grow, couple, and eventually form galaxies, stars, and planets. 

This perspective highlights why the inflationary epoch is so central to our cross-scale narrative. Everything that follows---from the cosmic web to biochemistry and brains---unfolds within a single realization of the stochastic ensemble produced at the end of inflation. The ``initial'' conditions for galaxy formation are not arbitrary; they are constrained by the requirement that they arise from a Gaussian field with specific correlations. In later sections we will see how gravitational instability maps these linear perturbations into nonlinear structures, effectively implementing a deterministic but highly sensitive dynamical evolution on the space of density fields.

\begin{figure}[t]
  \centering
  \includegraphics[width=\textwidth]{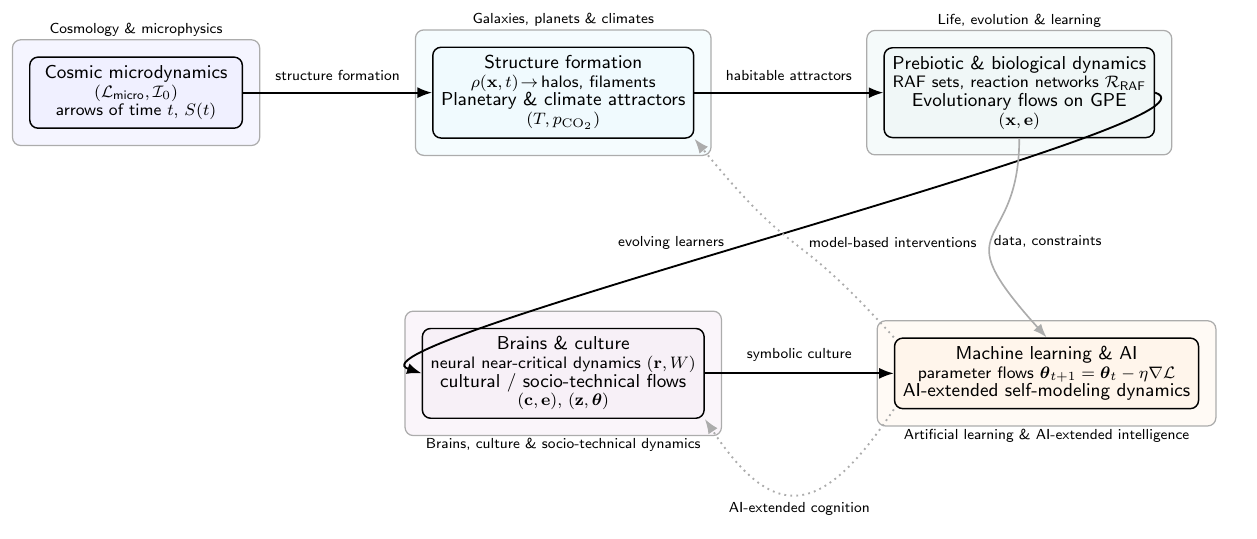}
  \caption{%
    Cross–scale dynamical narrative of the universe, from microphysics and cosmology
    (top-left) through structure formation and planetary attractors, to life, brains, culture,
    and machine learning / AI (bottom-right). Each block denotes an effective dynamical regime
    with its own natural state variables and flows; arrows indicate how later regimes
    emerge as constrained dynamics on the spaces created by earlier ones. The `Brains,
    culture \& socio–technical dynamics' and `Machine learning \& AI' blocks are
    unpacked in more detail in Figs.~\ref{fig:brains-culture-zoom}
    and~\ref{fig:ai-zoom}.}
  \label{fig:cosmic-overview}
\end{figure}

\subsection{Beyond the simplest picture: global structure and the measure problem}

The discussion above has focused on the simplest, single-patch view of inflation: a finite period of accelerated expansion that ends everywhere and leaves behind a universe like ours. Many modern models, however, suggest richer global structures. In ``chaotic'' and ``eternal'' inflation scenarios, different regions of the universe undergo varying amounts of inflation, perhaps exploring different minima of a complicated potential landscape \cite{Linde1986,Linde1984}. The global state space then includes not only the local degrees of freedom within a Hubble patch, but also the distribution of vacua, bubble nucleations, and domain structures on scales far beyond our observable horizon.

From a dynamical-systems standpoint, this raises the famously thorny \emph{measure problem}: how should we assign probabilities to different types of cosmological histories when the global space-time is infinite and contains infinitely many realizations of each local configuration? Different choices of time slicing, coarse-graining, or regularization can lead to different answers, and no consensus has emerged on a unique, ``natural'' measure \cite{Mukhanov2005}. For the purposes of this paper, we do not attempt to resolve these foundational questions. Instead, we adopt a pragmatic stance: we take as given that our observable patch is well-described by a single realization drawn from a Gaussian ensemble with parameters consistent with observations, and we investigate how the dynamics of later epochs sculpt that realization into structures of increasing complexity.

This modest stance is enough for our purposes because the later stages of our narrative---galaxy formation, planetary dynamics, biochemistry, evolution, and cognition---depend only on the local realization of perturbations within our Hubble volume, not on the global structure of the multiverse. Nevertheless, the inflationary measure problem serves as an instructive example of the general theme of this paper: whenever dynamics generate an ensemble of effective initial conditions, the question of which trajectories are realized and how typical they are cannot be answered without specifying both the microscopic laws and the appropriate coarse-grained measure on state space.

\smallskip

Ergo, the inflationary epoch is where our dynamical story truly begins. It takes a pre-inflationary universe whose detailed microstate we may never know, collapses its effective state space onto a slow-roll attractor, and populates that attractor with a structured ensemble of perturbations. The rest of cosmic history can then be read as the evolution of this ensemble under gravity, microphysics, and, eventually, biological and cognitive dynamics. In the next section, we turn to the first major act in this evolution: the growth of linear perturbations and their nonlinear collapse into the cosmic web of galaxies and clusters.

\section{From Linear Perturbations to the Cosmic Web: Gravitational Instability as Flow}
\label{sec:cosmic_web}

In the previous section, inflation handed us a nearly homogeneous universe with small, correlated density perturbations described by a Gaussian random field. From a dynamical-systems perspective, this is a point (plus small fluctuations) in an enormous state space whose coordinates are the density and velocity fields at each comoving position. The central claim of the standard structure-formation paradigm is that \emph{gravity alone}, acting on this initial condition, drives a deterministic flow on this function space that sculpts the familiar ``cosmic web'': voids, sheets, filaments, and clusters \cite{Peebles1980,Coles2001}. In this section we unpack that flow, starting from the linear regime and following it deep into the nonlinear patterns that define the backbone of the observable universe.

\subsection{The linear regime as flow in function space}

Let $\delta(\mathbf{x},t) \equiv [\rho(\mathbf{x},t)-\bar\rho(t)]/\bar\rho(t)$ denote the matter overdensity field in comoving coordinates, and let $\mathbf{v}(\mathbf{x},t)$ be the peculiar velocity. On sufficiently large scales and for cold dark matter, the evolution of $(\delta,\mathbf{v})$ in an expanding universe is governed by the continuity, Euler, and Poisson equations,
\begin{align}
\dot{\delta} + \frac{1}{a}\nabla\cdot\big[(1+\delta)\mathbf{v}\big] &= 0, \\
\dot{\mathbf{v}} + H\mathbf{v} + \frac{1}{a}(\mathbf{v}\cdot\nabla)\mathbf{v} &= -\frac{1}{a}\nabla\phi, \\
\nabla^2\phi &= 4\pi G a^2 \bar\rho\,\delta,
\end{align}
where $a(t)$ is the scale factor and $H=\dot{a}/a$ the Hubble parameter \cite{Peebles1980}. These equations define a flow on an infinite-dimensional state space: at each time $t$, the state is a point in the space of allowed $(\delta,\mathbf{v})$ fields, and the right-hand sides specify a vector field on that space.

Linearizing around $\delta=0$ and small velocities, we neglect quadratic terms to obtain
\begin{align}
\dot{\delta} + \frac{1}{a}\nabla\cdot\mathbf{v} &= 0,\qquad
\dot{\mathbf{v}} + H\mathbf{v} = -\frac{1}{a}\nabla\phi,
\end{align}
with the same Poisson equation. Combining these yields a second-order equation for $\delta$,
\begin{equation}
\ddot{\delta} + 2H\dot{\delta} - 4\pi G\bar\rho\,\delta = 0,
\label{eq:linear_growth}
\end{equation}
whose growing-mode solution defines a (scale-independent) growth factor $D(t)$ in standard $\Lambda$CDM cosmologies, such that in Fourier space
\begin{equation}
\delta(\mathbf{k},t) = D(t)\,\delta(\mathbf{k},t_\ast),
\end{equation}
for all modes well above the Jeans scale \cite{Peebles1980,SahniColes1995}. 

Equation~\eqref{eq:linear_growth} encapsulates gravitational instability as a linear, autonomous flow on the subspace of perturbations: each Fourier mode evolves independently with a growth rate set by the background cosmology. The Gaussian field produced by inflation thus simply ``rides'' this growth factor: if the initial field is Gaussian with known power spectrum $P(k)$, the field at later times remains Gaussian with amplified power $P(k,t)\propto D^2(t)P(k)$. In dynamical-systems terms, we have a linear semigroup acting on the space of perturbation fields; the early universe is in the basin of attraction of a homogeneous solution that is unstable to arbitrarily small perturbations.

A particularly illuminating way to view this is to imagine, at some early time, the universe as a high-dimensional point cloud: each point represents one realization of the initial Gaussian field drawn from the inflationary ensemble. Linear gravitational evolution stretches this cloud along a few unstable directions (associated with the growing mode) while contracting along decaying modes. The mapping remains one-to-one and volume-preserving in the space of random realizations, but the fields themselves are being amplified in physical space. Linear theory is thus the stage where we can still speak of a simple operator acting on the initial state; the complexity of the cosmic web will emerge when these flows drive many modes into the nonlinear regime and couple them together.

\subsection{Lagrangian dynamics and the Zel'dovich approximation}

The Eulerian picture above emphasizes the evolution of fields at fixed comoving positions. A complementary, and dynamically more intuitive, view follows individual fluid elements in \emph{Lagrangian} coordinates. Let $\mathbf{q}$ denote the initial (Lagrangian) position of a mass element and $\mathbf{x}(\mathbf{q},t)$ its comoving position at time $t$. In the Zel'dovich approximation \cite{Zeldovich1970,Shandarin2009}, one postulates that the displacement is proportional to the linear growth factor and the gradient of a potential:
\begin{equation}
\mathbf{x}(\mathbf{q},t) = \mathbf{q} - D(t)\,\nabla_{\mathbf{q}}\Phi(\mathbf{q}),
\label{eq:zeldovich}
\end{equation}
where $\Phi$ encodes the initial velocity potential. Mass conservation then implies that the density is given by
\begin{equation}
1 + \delta(\mathbf{x},t) = \left[\det\left(\frac{\partial \mathbf{x}}{\partial \mathbf{q}}\right)\right]^{-1}.
\end{equation}
Inserting \eqref{eq:zeldovich} reveals that overdensities grow where the deformation tensor has positive eigenvalues, with the collapse proceeding first along the direction of the largest eigenvalue, then the second, and finally the third.

This simple Lagrangian flow captures an essential feature of gravitational instability: it is intrinsically \emph{anisotropic}. Regions with one large eigenvalue collapse into \emph{pancakes} (sheets); with two large eigenvalues, into \emph{filaments}; and with three, into \emph{nodes} or halos. The Zel'dovich approximation thus provides a dynamical mapping from an initially featureless Gaussian random field to a network of lower-dimensional structures. Remarkably, it already reproduces the broad morphology of the cosmic web---walls, filaments, and compact clumps separated by expanding voids---well into the mildly nonlinear regime, until ``shell crossing'' (multi-streaming) occurs and the approximation breaks down \cite{SahniColes1995,Hidding2014}.

From the perspective of this paper, the key point is that \eqref{eq:zeldovich} is not just a heuristic trick; it is a first-order Lagrangian solution of the full gravitational dynamics and can be seen as a concrete instantiation of flow on a function space of mappings $\mathbf{q}\mapsto \mathbf{x}(\mathbf{q},t)$ \cite{Sahni1997}. The dynamical system here lives on the space of diffeomorphisms of comoving space, constrained by the requirement that these maps solve the Euler--Poisson system to a given order. The appearance of caustics and multi-stream regions when $\det(\partial \mathbf{x}/\partial \mathbf{q})\to 0$ signals the formation of singular structures---pancakes, filaments, and nodes---in a finite time. In this sense, the cosmic web is literally a network of singularities in the Lagrangian flow.

Extensions such as the adhesion model add an effective viscosity term that prevents trajectories from crossing and leads to ``sticking'' at caustics, sharpening filaments and nodes and yielding an even closer match to N-body simulations \cite{Sahni1997,Cautun2014}. These modifications can be viewed as adding a small amount of dissipation to an otherwise Hamiltonian system, causing the flow to concentrate measure onto a skeleton of lower-dimensional attractors.

\subsection{Nonlinear clustering and the emergence of the cosmic web}

As perturbations grow, the assumption of small $\delta$ fails on progressively larger scales, and the full nonlinearity of the gravitational equations becomes important. At this stage, analytic approaches like higher-order perturbation theory and Lagrangian expansions provide insight but must be complemented by large N-body simulations \cite{Peebles1980,Coles2001}. These simulations vividly reveal the web-like network of voids, sheets, filaments, and clusters that defines the large-scale structure of the universe.

Bond, Kofman, and Pogosyan famously interpreted this pattern as the natural outcome of anisotropic gravitational collapse in a Gaussian random field: the initial tidal field determines the directions of fastest collapse, leading to a hierarchy in which matter first forms sheets, then filaments at their intersections, and finally dense knots at filament junctions \cite{BondKofmanPogosyan1996,BondWebReview}. Subsequent work has sharpened this picture, using dynamical classifications based on the eigenvalues of the deformation tensor or the velocity shear to label regions of a simulation as voids, walls, filaments, or nodes and to track their evolution over time \cite{Cautun2014,Demianski1999}.

In our language, we can think of the nonlinear regime as the point where the flow in the state space of density fields becomes strongly mixing: different Fourier modes couple, the Gaussian probability measure is pushed forward into a highly non-Gaussian one, and the dynamics generate new effective degrees of freedom---the positions and internal structures of halos, the connectivity of filaments, and the morphology of voids. Nevertheless, the ghost of the initial conditions persists: the large-scale skeleton of the web, its primary filaments and voids, can be traced back directly to the initial Gaussian field via the Zel'dovich mapping \cite{Hidding2014}. The cosmic web is thus a dynamically evolved but still recognizable image of the inflationary random field, processed by gravity.

Statistically, measures such as the two-point correlation function, power spectrum, Minkowski functionals, and genus statistics quantify how far the evolved density field has moved from its initial Gaussian state \cite{Kerscher1999,Sahni1999}. These descriptors can be understood as coarse probes of the underlying flow: for example, the growth of higher-order moments and topological descriptors directly tracks the nonlinear coupling of modes and the formation of connected structures. In this way, large-scale structure surveys become experiments in dynamical systems: by observing the present-day configuration of galaxies, we infer properties of the gravitational flow and the initial measure.

\subsection{Halos as quasi-attractors and hierarchical assembly}

On smaller scales, the nonlinear flow fragments the cosmic web into a hierarchy of bound dark matter halos---the sites where galaxies and clusters form. A widely used analytic description of this process is the Press--Schechter formalism, which treats collapse as occurring when the linearly extrapolated density contrast in a region, smoothed on a given mass scale, exceeds a critical threshold $\delta_c$ corresponding to spherical collapse \cite{PressSchechter1974,PressSchechterReview}. Under the assumption of Gaussian initial conditions, this yields an explicit prediction for the mass function of collapsed halos as a function of time. 

From a dynamical viewpoint, halos play the role of \emph{quasi-attractors}: once matter falls into a sufficiently deep potential well and virializes, it executes long-lived, bound orbits. While the full N-body system remains Hamiltonian and has no true attractors in the strict sense, coarse-graining over phase space shows that mass flows into a relatively small subset of configurations (bound halos) and remains there for cosmologically long times \cite{ColeLacey1995,MoWhite1996}. Continuous infall along filaments and mergers between halos then implement a hierarchical assembly process in which small structures form first and subsequently merge into larger ones.

This hierarchical clustering introduces a new emergent description: instead of tracking the full density field, we can describe the state of the system by a point process of halos with masses, positions, velocities, and internal density profiles. In other words, the gravitational flow on the original function space induces a derived flow on a reduced state space of ``objects'' and their relations. This kind of effective, coarse-grained state space will reappear when we discuss biological and cultural evolution: just as galaxies and clusters become the relevant degrees of freedom for cosmology at late times, organisms and institutions become the relevant degrees of freedom for evolution and history.

\smallskip

Ergo, gravitational instability transforms the nearly featureless Gaussian field produced by inflation into the richly structured cosmic web via a sequence of dynamical stages: linear amplification of modes, anisotropic collapse captured by the Zel'dovich approximation, nonlinear mode coupling, and hierarchical formation of quasi-bound halos. At each stage, the flow in the state space of density fields amplifies particular directions, forms singular structures, and effectively reduces the dimensionality of the dynamics by concentrating matter into a network of lower-dimensional features. In the next section, we zoom in on one branch of this web: the dissipative collapse of gas within dark-matter halos, and the emergence of stars and planetary systems as new dynamical attractors riding on top of the collisionless skeleton.

 \section{Dissipative Collapse and Stellar Attractors: Stars, Planets, and Nucleosynthesis}
\label{sec:stellar_planetary}

The cosmic web we described in the previous section is built from collisionless dark matter. By itself, this web would remain a cold, sparse scaffold of halos and filaments. The luminous universe we actually observe---stars, nebulae, planets, and the chemical ingredients for life---appears only when a dynamically distinct component, baryonic gas, is added to the picture. Unlike dark matter, gas can radiate, cool, shock, fragment, and form bound thermonuclear engines. In dynamical-systems terms, baryons introduce new dissipative degrees of freedom that allow trajectories to flow from large-scale, nearly Hamiltonian orbits in halos down to small-scale, long-lived attractors: stars, stellar remnants, and planetary systems. This section follows that cascade.

\subsection{Baryons in dark-matter wells: cooling and angular-momentum funnels}

Once dark-matter halos have formed via gravitational instability, baryonic gas falls into their potential wells and is shock-heated to approximately the virial temperature,
\begin{equation}
k_B T_{\rm vir} \sim \frac{\mu m_p}{2} V_c^2,
\end{equation}
where $V_c$ is the circular velocity and $\mu m_p$ the mean particle mass. In the absence of cooling, this gas would simply remain in quasi-hydrostatic equilibrium, tracing the dark-matter potential and never condensing further. The crucial dynamical ingredient is radiative cooling: atomic lines, free--free emission, and molecular transitions provide channels for the gas to lose internal energy and entropy, contracting toward denser configurations \cite{WhiteRees1978}.

The White--Rees ``two-stage'' picture of galaxy formation captures this structure succinctly: collisionless dark matter builds the halos; dissipative baryons cool, condense, and fragment within those halos, forming the luminous cores we call galaxies \cite{WhiteRees1978}. The relevant control parameter is the ratio of the cooling time to the dynamical (free-fall) time, $t_{\rm cool}/t_{\rm dyn}$. Regions where $t_{\rm cool} \ll t_{\rm dyn}$ experience runaway collapse: as gas contracts, it heats but radiates the energy away even faster, sliding down a steep potential in thermodynamic state space. Where $t_{\rm cool} \gg t_{\rm dyn}$, the gas is effectively ``stuck'' in a hot, pressure-supported quasi-equilibrium.

Angular momentum adds another layer of geometry. Infalling gas generically carries nonzero specific angular momentum, inherited from tidal torques during halo assembly. Conservation of angular momentum prevents collapse to a point; instead, gas settles into rotationally supported, roughly axisymmetric discs. From a dynamical-systems perspective, this is a flow from a three-dimensional family of orbits into a lower-dimensional manifold: a disc with a characteristic surface density and scale height, slowly evolving under viscosity, magnetic torques, and self-gravity. The large-scale state space of baryons in a halo thus contains a quasi-attractor corresponding to a centrifugally supported disc, whose subsequent fragmentation seeds star formation \cite{WhiteRees1978,Krumholz2014}.

\subsection{Stars as thermonuclear attractors in stellar-structure space}

Within cold, dense regions of galactic discs---giant molecular clouds---self-gravity, turbulence, magnetic fields, and feedback from young stars conspire to produce a spectrum of bound condensations that we call prestellar cores \cite{ShuAdamsLizano1987,Krumholz2014}. On scales of individual cores, the dynamics simplify dramatically: to good approximation, a single, self-gravitating fluid sphere cools, contracts, and eventually becomes opaque. Once the interior becomes sufficiently hot and dense, thermonuclear reactions ignite, and the object adjusts toward a state of hydrostatic and thermal equilibrium: a star.

Mathematically, the interior of a non-rotating star is described by a set of coupled first-order differential equations for mass, pressure, temperature, and luminosity as functions of radius \cite{KippenhahnWeigert1990,Prialnik2009}. For given boundary conditions and composition, these equations define a family of equilibrium solutions parameterized primarily by mass. In the $(\log T_{\rm eff}, \log L)$ plane of the Hertzsprung--Russell diagram, these equilibria lie along the familiar main sequence and subsequent evolutionary tracks \cite{KippenhahnWeigert1990}. 

From a dynamical-systems viewpoint, these equilibria are not static curiosities; they are \emph{attractors} for the stellar evolution flow on a higher-dimensional state space that includes internal composition profiles and entropy. On short dynamical time scales, the star rapidly relaxes to hydrostatic equilibrium via sound waves, extinguishing large-scale fluid motions. On intermediate Kelvin--Helmholtz time scales, radiative diffusion and convection rearrange entropy until the internal luminosity profile matches what nuclear reactions can support. On the longest nuclear time scales, composition variables drift slowly as fuel is burned. The separation of these time scales allows us to treat a star as executing a slow trajectory along a manifold of quasi-equilibrium configurations: a one-dimensional curve in a high-dimensional phase space.

Nuclear reaction networks drive this slow drift. In the simplest picture, the abundances of nuclei $\{Y_i\}$ evolve according to a stiff system of ordinary differential equations,
\begin{equation}
\dot{Y}_i = \sum_{j,k} N_{ijk}\,\lambda_{jk}(\rho,T)\,Y_j Y_k + \cdots,
\end{equation}
where $\lambda_{jk}$ are temperature- and density-dependent reaction rates and $N_{ijk}$ encode stoichiometric coefficients. The landmark B$^2$FH paper showed how different combinations of such reactions---proton–proton chains, CNO cycles, $\alpha$-captures, and neutron-capture processes---populate the periodic table inside stars and during their explosive deaths \cite{Burbidge1957}. In the space of composition vectors $\mathbf{Y}$, these networks define a vector field whose integral curves describe nucleosynthetic pathways. The star’s macroscopic structure responds quasi-statically to the slowly changing $\mathbf{Y}$, moving through a sequence of core and shell-burning phases.

Crucially, different initial masses and metallicities land the system in qualitatively different basins of attraction. Low-mass stars ($M \lesssim 8\,M_\odot$) evolve toward white-dwarf remnants, losing their envelopes and leaving behind degenerate carbon–oxygen or oxygen–neon cores. More massive stars traverse a series of central burning stages up to iron, approach regions of dynamical instability, and then undergo core collapse, producing neutron stars or black holes and ejecting heavy elements in supernova explosions \cite{KippenhahnWeigert1990,Burbidge1957}. Each branch corresponds to a distinct attractor class in the enlarged state space that includes possible compact remnant configurations and their surrounding ejecta.

\subsection{Chemical enrichment as flow in abundance space}

Viewed on galactic scales, individual stellar life cycles become nodes in a larger reaction network: the interstellar medium (ISM) is continually enriched and stirred by stellar feedback. Each generation of star formation samples the ISM composition, evolves along its stellar-structure attractors, and returns processed material via winds and supernovae. Over cosmic time, this defines a coarse-grained flow on a lower-dimensional state space spanned by bulk metallicity $Z$ and selected abundance ratios (e.g., $\alpha$/Fe, C/O).

Simple ``closed-box'' and leaky-box models of galactic chemical evolution treat the gas mass $M_{\rm gas}$ and metallicity $Z$ as dynamical variables obeying
\begin{align}
\dot{M}_{\rm gas} &= -\psi(t) + E(t) - O(t), \\
\dot{(Z M_{\rm gas})} &= -Z\,\psi(t) + E_Z(t) - Z\,O(t),
\end{align}
where $\psi$ is the star-formation rate, $E$ and $E_Z$ are total mass and metal return from stellar ejecta, and $O$ represents outflows \cite{Burbidge1957,Wallerstein1997}. Although highly idealized, such models capture an important dynamical fact: repeated cycling through stellar attractors pushes the ISM along characteristic trajectories in abundance space, with the details depending on the stellar initial mass function (IMF) and feedback physics.

Empirically, the IMF appears to be nearly universal across many environments, with a broken power-law or lognormal form summarized by the Kroupa and Chabrier parameterizations \cite{Kroupa2001,Chabrier2003}. This near-universality implies that the mapping from a given star-formation history to chemical-evolution trajectories is relatively robust. Ergo, the IMF defines a coarse-grained measure on the space of stellar evolution paths, and the convolution of that measure with nucleosynthetic yields defines the effective vector field driving chemical evolution. The upshot is that the ISM is not wandering randomly through abundance space; it is being drawn toward patterns set by the interplay of star formation, feedback, and gas flows.

\subsection{Protoplanetary discs and planet formation as disc instabilities}

The same dissipative processes that allowed baryons to condense into stars also create the environments where planets form. Young stars are typically surrounded by gas-rich protoplanetary discs that act as angular-momentum reservoirs and cradles for solid-body growth \cite{Armitage2011}. These discs are quasi-steady accretion flows: viscous and magnetic stresses transport angular momentum outward, allowing mass to drift inward and accrete onto the star. In a dynamical-systems picture, the disc is a slowly evolving, nearly two-dimensional fluid configuration, susceptible to a range of linear and nonlinear instabilities.

Dust grains embedded in the disc experience gas drag, causing them to drift radially and settle toward the midplane. As solids become locally concentrated, the coupled gas–dust system can develop the ``streaming instability'', a drag-driven linear instability where relative streaming of particles and gas leads to exponential growth of particle overdensities \cite{YoudinGoodman2005,Squire2020}. Numerical simulations show that this instability can concentrate solids by orders of magnitude, triggering gravitational collapse into planetesimals. Here again, we see a familiar motif: a nearly homogeneous state (well-mixed dust and gas) is linearly unstable, and the ensuing flow funnels measure in state space onto lower-dimensional structures (dense clumps) that act as seeds for further hierarchical assembly.

Classical planetesimal formation models, going back to Safronov's work on the evolution of the protoplanetary cloud, treat the growth of solid bodies as a coagulation process driven by inelastic collisions and gravitational focusing \cite{Safronov1969}. In such models, the mass distribution of solids evolves according to kinetic equations reminiscent of Smoluchowski coagulation, with solutions that can exhibit runaway growth of a few large bodies embedded in a sea of small ones. More recent theories combine coagulation with streaming instabilities, pebble accretion, and migration to explain the rich diversity of observed exoplanetary systems, from hot Jupiters to compact resonant chains \cite{Armitage2011,Johansen2014}.

Ergo, the emergence of stars and planets can be read as a series of dissipative cascades riding on the collisionless cosmic web: gas cools and collapses within dark-matter halos; stars arise as thermonuclear attractors in stellar-structure space; repeated stellar cycles drive a flow through chemical-abundance space; and protoplanetary discs fragment and coagulate into planetary systems. Each stage implements the same basic dynamical story: radiative and collisional processes provide a way for matter to shed energy and entropy, contracting the effective phase space and populating a hierarchy of long-lived, structured attractors. In the next section, we shift from astrophysical to planetary scales and examine Earth-like planets as multiscale dynamical systems, where climate, geology, and geochemistry conspire to maintain habitable, far-from-equilibrium states for billions of years.

\section{Planets as Multiscale Dynamical Systems: Climate, Geology, and Cycles}
\label{sec:planetary_dynamics}

Dark-matter halos and stellar attractors provide the scaffolding and energy sources for habitable environments, but they do not guarantee them. Whether a given rocky world becomes a frozen snowball, a Venus-like hothouse, or a long-lived, life-bearing planet depends on the interplay of climate, geology, and geochemistry on scales from turbulent eddies to mantle overturn and continental drift. In this section we treat Earth-like planets as \emph{multiscale dynamical systems}: coupled subsystems with different time scales and feedbacks, whose interaction generates multiple climate attractors, hysteresis, and the possibility of self-stabilization or runaway. This perspective links classic work on climate dynamics and paleoclimate \cite{Lorenz1963,Saltzman2002}, long-term carbon cycling \cite{WalkerHaysKasting1981}, and Earth system feedbacks \cite{LovelockMargulis1974,Lenton2016} with modern ideas about planetary habitability and tipping points \cite{Kasting1993,Lenton2000,Arnscheidt2022}.

\subsection{A minimal dynamical skeleton: energy balance and feedbacks}

At the most aggregated level, a planet’s climate can be modeled by a global mean energy-balance equation,
\begin{equation}
C\,\dot{T} = \frac{S(1-\alpha(T,\ldots))}{4} - \sigma T^4 + F_{\rm greenhouse}(T, p_{\mathrm{CO}_2},\ldots),
\label{eq:ebm}
\end{equation}
where $T$ is a characteristic surface temperature, $C$ an effective heat capacity, $S$ the stellar constant, $\alpha$ the planetary albedo, and $F_{\rm greenhouse}$ aggregates infrared back-radiation from greenhouse gases and clouds \cite{Kasting1993}. Despite its simplicity, \eqref{eq:ebm} already defines a one-dimensional dynamical system with feedbacks: $\alpha$ and $F_{\rm greenhouse}$ depend on $T$ and other variables, encoding ice–albedo feedbacks (colder $\to$ more ice $\to$ higher albedo $\to$ colder) and water-vapour and cloud feedbacks (warmer $\to$ more greenhouse effect $\to$ warmer).

More realistic climate models add spatial structure and fluid dynamics, yielding high-dimensional systems with rich behaviour. Lorenz’s celebrated three-variable model of convection,
\begin{equation}
\dot{X} = \sigma(Y-X),\quad
\dot{Y} = rX - Y - XZ,\quad
\dot{Z} = XY - bZ,
\end{equation}
was originally derived as a low-order truncation of the equations for thermal convection in a fluid layer \cite{Lorenz1963}. It already exhibits sensitive dependence on initial conditions and a strange attractor: a compact, fractal subset of state space on which trajectories wander chaotically. Saltzman’s work on dynamical paleoclimatology generalizes this approach, embedding ice volume, greenhouse gases, and ocean circulation into coupled nonlinear models that reproduce glacial cycles, millennial variability, and shifts between climate regimes \cite{Saltzman2002}. In both cases, climate is not a static “equilibrium” but an evolving trajectory on a structured attractor with multiple time scales.

From our cross-scale perspective, these models serve two roles. First, they demonstrate that even very low-dimensional truncations of planetary climate can support multiple equilibria, limit cycles, and chaotic attractors, depending on parameters. Second, they illustrate how fast atmospheric processes and slower cryospheric or biogeochemical processes couple through feedbacks to create new effective variables: for example, ice-line latitude or deep-ocean temperature, which evolve slowly but shape the geometry of the attractor.

\subsection{The long-term carbon–silicate thermostat}

On geological time scales ($\gtrsim 10^6$\,yr), radiative and dynamical feedbacks are not enough to explain the relative stability of Earth’s surface temperature in the face of a steadily brightening Sun. A classic proposal by Walker, Hays, and Kasting \cite{WalkerHaysKasting1981} introduces a chemical negative feedback: the \emph{carbonate–silicate cycle}. In their picture, atmospheric $\mathrm{CO}_2$ is removed by weathering of continental silicate rocks and subsequent precipitation of carbonates in the oceans, while it is added by volcanic and metamorphic degassing. Crucially, the weathering rate increases with temperature, precipitation, and (to a lesser extent) $p_{\mathrm{CO}_2}$, providing a stabilizing loop: a warmer climate accelerates weathering, which draws down $\mathrm{CO}_2$ and cools the planet; a cooler climate slows weathering, allowing $\mathrm{CO}_2$ to accumulate and warm the surface.

In dynamical form, one can write a minimal system,
\begin{align}
\dot{T} &= f(T, p_{\mathrm{CO}_2}, S,\ldots), \\
\dot{p}_{\mathrm{CO}_2} &= F_{\rm outgassing} - F_{\rm weather}(T,p_{\mathrm{CO}_2},\ldots),
\end{align}
where $f$ encodes radiative balance as in \eqref{eq:ebm} and $F_{\rm weather}$ increases steeply with $T$ \cite{WalkerHaysKasting1981,Arnscheidt2022}. The fixed points of this system correspond to long-term climate equilibria where outgassing balances weathering and radiative input balances output. Linearizing around such a fixed point shows that, for reasonable parameter choices, the thermostat is stable: small perturbations in $T$ and $p_{\mathrm{CO}_2}$ decay on weathering time scales, even as solar luminosity slowly increases.

Recent work has revisited and generalized this mechanism, exploring the conditions under which Earth-like planets possess stabilizing weathering feedbacks versus cases where weathering saturates or becomes supply-limited \cite{Arnscheidt2022}. The geometry of the resulting vector field in $(T,p_{\mathrm{CO}_2})$ space determines whether trajectories slowly track a moving equilibrium (a robust thermostat) or veer toward runaway greenhouse or snowball states when external forcing crosses certain thresholds. In other words, the long-term habitability of a planet is a question about the existence and stability of attractors in this reduced state space, and about how they bifurcate as stellar flux and tectonic or biological parameters vary.

\subsection{Multiple climate attractors: Snowball Earth and hysteresis}

The same feedbacks that stabilize climate near a temperate state can also create multiple attractors separated by unstable manifolds, allowing history and perturbations to matter in a strong sense. The \emph{Snowball Earth} hypothesis, first articulated in modern form by Kirschvink \cite{Kirschvink1992} and developed by Hoffman and co-workers \cite{Hoffman1998,Hoffman2002}, proposes that during the Neoproterozoic, Earth experienced episodes of near-global glaciation, with ice reaching low latitudes. Geological evidence (e.g., low-latitude glacial deposits, cap carbonates, and isotopic anomalies) suggests that these snowball states were entered and exited relatively abruptly \cite{Hoffman1998,Hoffman2002}.

In a dynamical-systems picture, Snowball Earth corresponds to a separate climate attractor characterized by high albedo and low $T$, distinct from the present temperate attractor. Energy-balance models with temperature-dependent albedo generically exhibit such bistability: for certain ranges of stellar forcing and greenhouse gas concentration, the system admits both a warm solution with small ice caps and a cold solution with nearly global ice cover, separated by an unstable intermediate state. Trajectories crossing the unstable manifold (for example, due to gradual changes in $S$, continental configuration, or $p_{\mathrm{CO}_2}$) can tip from one attractor to the other, exhibiting hysteresis when the forcing is reversed \cite{Kirschvink1992,Hoffman2002,Saltzman2002}.

Deglaciation from a snowball state likely required extremely high $\mathrm{CO}_2$ levels built up by continued volcanic outgassing while silicate weathering was nearly shut down under ice, a scenario consistent with the carbonate–silicate thermostat \cite{Hoffman1998,Kirschvink1992}. Once $p_{\mathrm{CO}_2}$ exceeded a critical threshold, radiative forcing would have overwhelmed the ice–albedo feedback, driving a catastrophic melt and a transition back to the warm attractor. Here we see clearly how coupled geochemical and radiative feedbacks generate a nontrivial topology of climate attractors and basins in $(T,p_{\mathrm{CO}_2})$ space---and how geological events correspond to heteroclinic transitions between them.

\subsection{Earth system feedbacks and biosphere–climate coupling}

So far we have treated climate and geochemistry as a purely physical–chemical system. On a living planet, the biosphere introduces additional feedbacks. The Gaia hypothesis of Lovelock and Margulis \cite{LovelockMargulis1974} proposed that life and environment co-evolve to maintain habitable conditions through a network of feedbacks. Early formulations were criticized for teleology, but stripped of purpose-driven language, the core claim is dynamical: biological processes can create stabilizing (or destabilizing) feedback loops that alter the location and stability of climate attractors.

Lovelock’s ``Daisyworld'' toy model illustrates this idea in perhaps its purest dynamical form: a simple planet with black and white daisies whose growth depends on local temperature, and whose albedos modify the climate. The coupled system self-organizes so that, over a range of stellar fluxes, the planetary temperature remains near the daisies’ optimum, even though neither local nor global regulation is explicitly programmed \cite{LovelockMargulis1974,Zeng1990}. In state-space terms, the interaction between biota and climate reshapes the vector field so that a robust attractor with small temperature variations appears over a wider parameter range than in the abiotic case.

Modern Earth system science reframes this more quantitatively. Lenton and colleagues have catalogued a suite of feedbacks linking climate to biogeochemical cycles and ecosystems, from land and ocean carbon uptake to vegetation–albedo coupling and cloud–aerosol interactions \cite{Lenton2000,Lenton2016}. Simple models and comprehensive Earth system models show how these feedbacks can either stabilize climate (negative feedbacks) or create the possibility of tipping points when thresholds are crossed (positive feedbacks). In dynamical-systems language, biological processes modify both the locations of fixed points and their eigenvalue spectra, altering the number and stability of climate attractors.

Recent work has asked, more generally, under what conditions Earth-like planets possess ``Gaian'' stabilizing feedbacks as opposed to fragile or even self-destructive (``Medean'') feedbacks \cite{Arnscheidt2022}. The answer depends on the co-evolution of life and environment, and can be phrased as a problem in evolutionary dynamics on a coupled climate–biosphere state space: which feedback-producing traits are favoured, and what dynamical regimes do they induce? Although we lack a general theory, the conceptual point is clear: habitability is not just a property of external forcing and initial conditions, but of the feedback architecture---the way fast and slow subsystems couple and reshape each other’s attractor structure.

\subsection{Beyond Earth: habitable zones as regions in parameter space}

Finally, from the broader cosmic perspective, Earth is one point in a much larger parameter space of possible planetary dynamical systems. Kasting’s classical habitable-zone calculations treat the inner and outer edges of habitability as bifurcation curves in $(S, a)$ space (stellar flux and orbital distance), where attractors corresponding to temperate, water-bearing climates cease to exist or become unstable due to runaway greenhouse or CO$_2$ condensation \cite{Kasting1993}. More recent work extends this to three-dimensional climate models and a range of planetary properties, including obliquity, rotation rate, and continental configuration \cite{Williams1997,Godolt2016}.

From our standpoint, the habitable zone is best seen as a region in parameter space where the coupled climate–geology–biogeochemistry dynamical system admits at least one long-lived, structurally stable attractor with surface conditions compatible with liquid water and complex chemistry. The precise shape of this region depends sensitively on the presence and strength of feedbacks like the carbonate–silicate cycle and biosphere–climate coupling. In this sense, the existence of a robust, wide habitable zone is not simply an astrophysical question (stellar type and insolation) but a question about the internal dynamical architecture of planets.

\smallskip

Ergo, Earth-like planets are not passive recipients of stellar radiation and cosmic initial conditions; they are high-dimensional dynamical systems whose climate, geology, and biosphere co-evolve on nested time scales. Negative feedbacks like the carbonate–silicate cycle can act as thermostats, carving out stable attractors that track a brightening star, while positive feedbacks and nonlinearities create multiple attractors and tipping points, as in Snowball Earth episodes. Biological processes further reshape the flow, potentially widening or shrinking the parameter region supporting habitable attractors. In the next section, we zoom into one such habitable attractor---early Earth’s---and ask how, within its chemically active oceans and atmospheres, the dynamics of reaction networks gave rise to the first self-maintaining, living systems.

\section{From Prebiotic Chemistry to Autocatalytic Sets: The Origin of Living Dynamics}
\label{sec:origin_life}

Within a habitable planetary attractor, the story of life begins not with organisms, but with \emph{chemistry} that refuses to relax to equilibrium. Thin films, hydrothermal pores, or tidal pools host vast reaction networks driven by geothermal, redox, and photochemical free-energy sources. Out of this turbulence, a tiny subset of chemistries begin to do something qualitatively new: they maintain and reproduce their own organization. In this section we recast origin-of-life scenarios as questions about when and how reaction networks cross a dynamical threshold from mere \emph{geochemistry} to \emph{living dynamics}.

\subsection{Reaction networks as dynamical systems}

Prebiotic chemistry can be modeled as a set of molecular species $\{X_i\}$ participating in reactions
\[
\sum_i \nu^-_{i\rho} X_i \;\xrightleftharpoons[k_{-\rho}]{k_{\rho}}\; \sum_i \nu^+_{i\rho} X_i,
\]
where $\rho$ indexes reactions, $\nu^\pm_{i\rho}$ are stoichiometric coefficients, and $k_{\pm\rho}$ are rate constants. Under well-mixed conditions and mass-action kinetics, the concentration vector $\mathbf{x}(t)$ evolves according to
\begin{equation}
\dot{\mathbf{x}} = N\,\mathbf{v}(\mathbf{x}), \qquad
v_\rho(\mathbf{x}) = k_{\rho}\,\prod_i x_i^{\nu^-_{i\rho}} - k_{-\rho}\,\prod_i x_i^{\nu^+_{i\rho}},
\label{eq:crn}
\end{equation}
where $N$ is the stoichiometric matrix with entries $N_{i\rho} = \nu^+_{i\rho} - \nu^-_{i\rho}$.
Equation~\eqref{eq:crn} defines a flow on a high-dimensional concentration space, subject to linear conservation laws (e.g., total atoms of each element) that confine trajectories to stoichiometric compatibility classes. If the network is closed and detailed balance holds, this flow converges to a unique thermodynamic equilibrium.

However, prebiotic environments were almost certainly \emph{open}: small molecules were continuously supplied and waste products removed by geological fluxes, sunlight, or hydrothermal circulation. Mathematically, this means some species are effectively buffered (chemostatted) or coupled to boundary fluxes, breaking detailed balance and allowing nonequilibrium steady states and oscillations. In such open chemical reaction networks, the long-term behaviour is organized by invariant sets (steady states, cycles) and their basins, just as in any dynamical system. Origin-of-life theories can thus be framed as the search for network topologies and boundary conditions under which trajectories are funneled into self-maintaining, self-amplifying organizations rather than dissipating into featureless equilibrium \cite{Kauffman1993,Krishnamurthy2020}.

\subsection{Autocatalytic sets and RAF theory: self-maintaining subnetworks}

Kauffman’s proposal of collectively autocatalytic sets (CAS) gave a first precise articulation of this idea: a set of molecules and reactions is \emph{autocatalytic} if the molecules catalyze the very reactions that produce them, forming a closed loop of mutual support \cite{Kauffman1993}. More recently, this notion has been formalized as RAF theory (reflexively autocatalytic and food-generated sets) by Hordijk, Steel, and collaborators \cite{HordijkSteel2004,Hordijk2017Review,Steel2019}. In RAF language, a catalytic reaction system consists of:

\begin{itemize}
    \item a set of molecule types $\mathcal{X}$,
    \item a set of reactions $\mathcal{R}$ on $\mathcal{X}$,
    \item a catalysis relation specifying which molecules catalyze which reactions,
    \item and a ``food set'' $F \subset \mathcal{X}$ of simple molecules supplied by the environment.
\end{itemize}

A subset $\mathcal{R}' \subseteq \mathcal{R}$ is a RAF if (i) it is \emph{reflexively autocatalytic}: every reaction in $\mathcal{R}'$ is catalyzed by at least one molecule that is either in $F$ or produced by $\mathcal{R}'$, and (ii) it is \emph{food-generated}: all reactants in $\mathcal{R}'$ can be built from $F$ using only reactions in $\mathcal{R}'$ \cite{Hordijk2017Review,Steel2019}. Intuitively, a RAF is a self-sustaining subnetwork that can bootstrap itself from environmental food.

From a dynamical-systems viewpoint, RAFs identify candidate \emph{self-maintaining attractors} in the full reaction network. Even though the detailed kinetics \eqref{eq:crn} may be complex, the existence of a RAF guarantees that, once a minimal subset of its molecules is present above threshold, reaction fluxes can, in principle, maintain those molecules against dilution and decay, provided food and energy are supplied. Moreover, mathematical results show that RAFs arise generically once the density of catalysis crosses a critical threshold in random reaction networks, and that RAF-like structures can be detected in real biochemical systems, including core metabolism \cite{Hordijk2017Review,Steel2019,Hordijk2023}. This resonates with the themes of earlier sections: a large, high-dimensional dynamical system spontaneously develops a lower-dimensional, self-maintaining submanifold that captures much of its long-term behaviour.

The internal structure of RAFs also supports the emergence of modularity and hierarchy. Maximal RAFs can often be decomposed into smaller, irreducible RAFs linked in a partial order, suggesting a natural route by which more complex organizations could accrete from simpler autocatalytic motifs \cite{Hordijk2017Review}. In phase-space language, the full chemical flow may contain multiple nested or overlapping self-sustaining regions, each corresponding to a different chemical ``ecosystem''. The origin-of-life problem becomes, at least partly, the problem of understanding when such regions appear, how robust they are to perturbations and parasitic reactions, and how they can be localized and amplified.

\subsection{Templates, replicators, and the RNA world}

Autocatalytic sets provide a structural criterion for self-maintenance, but living systems also require \emph{heritable information}. One influential hypothesis is the RNA world: an early stage in which RNA molecules acted both as carriers of genetic information and as catalysts (ribozymes) \cite{Gilbert1986RNAworld,RobertsonJoyce2012}. In this picture, some subset of prebiotic chemistry is dominated by template-directed polymerization reactions in which an RNA strand $R_i$ catalyzes the production of copies (or variants) of itself from activated monomers. 

The dynamics of such replicators can be described, at a coarse-grained level, by replicator equations:
\begin{equation}
\dot{x}_i = x_i\bigl(f_i(\mathbf{x}) - \bar{f}(\mathbf{x})\bigr),
\end{equation}
where $x_i$ is the fraction of type $i$ and $f_i$ its replication rate, with $\bar{f}$ the population-average fitness \cite{EigenSchuster1979}. Eigen and Schuster’s hypercycle model is a particularly elegant example, in which different replicator species catalyze each other’s formation in a cyclic network, forming a higher-order autocatalytic organization \cite{EigenSchuster1979}. Hypercycles are attractors in the space of replicator frequencies; they can store more information collectively than any single species could withstand under mutation alone, but are also vulnerable to parasitic sequences.

The RNA world and RAF/autocatalytic perspectives are not mutually exclusive. An RNA-based protocell could host an autocatalytic set of ribozymes that collectively maintain a metabolic core, while template replication supplies heredity. RAF theory has, in fact, been applied to RNA networks and shows that self-sustaining, collectively autocatalytic RNA sets can emerge under plausible conditions \cite{Steel2019}. In this unified view, the key dynamical step is the appearance of a \emph{coupled} attractor: a set of reactions and polymers that both maintains its chemical organization and transmits enough structured variation to support Darwinian evolution.

\subsection{Protocells and compartments: from chemistry to individuals}

All of the above dynamics initially take place in extended environments: a rock pore, a pond, a vent system. The next qualitative leap is \emph{compartmentalization}. Protocell models posit small, self-assembled compartments—typically fatty-acid vesicles or other amphiphilic aggregates—that encapsulate subsets of the reaction network \cite{Morowitz1992Beginnings,SchrumSzostak2010,Szostak2017DeepPast}. These compartments:

\begin{itemize}
    \item create local chemical microenvironments with distinct compositions,
    \item selectively retain larger catalysts while exchanging small nutrients and wastes,
    \item and can grow and divide based on physical processes (osmotic swelling, shear, budding).
\end{itemize}

From a dynamical standpoint, compartments introduce a new level of description. Instead of a single concentration vector $\mathbf{x}(t)$ for the whole environment, we now have many vectors $\mathbf{x}^{(k)}(t)$, one for each protocell $k$, plus variables describing membrane properties and populations. Internal reaction networks (possibly autocatalytic, possibly including replicators) drive growth of the compartment, while the environment supplies food molecules and removes waste. When protocells grow and divide, the system effectively implements a stochastic map on the space of internal compositions: contents are partitioned between daughters, and slight asymmetries can be amplified over generations \cite{SchrumSzostak2010,Saha2015Protocell}.

This multi-level system naturally supports selection. Protocells whose internal networks draw more effectively on environmental resources will, on average, grow and divide faster, increasing their representation in the population. Internal chemistries that promote membrane growth or nutrient import are favoured, even if they do not self-replicate perfectly on their own. In other words, the dynamical attractors relevant to long-term behaviour shift: from fixed points or cycles in chemical concentration space to attractors in the space of protocell lineages, with internal chemistry acting as a high-dimensional parameterization of fitness.

\subsection{The phase transition to Darwinian evolution}

At what point, if any, should we say that a prebiotic system has crossed from ``mere chemistry'' into bona fide life? One operational answer is: when the appropriate dynamical description switches from reaction-network dynamics to evolutionary dynamics on a space of heritable types. That is, when the system supports:

\begin{enumerate}
    \item \emph{Reproduction}: entities (e.g., protocells) that can generate similar entities,
    \item \emph{Variation}: imperfect copying, generating a distribution over types,
    \item \emph{Differential persistence}: systematic differences in long-term growth or survival.
\end{enumerate}

Once these conditions hold, the relevant state variables are no longer only concentrations and fluxes, but also genotype (or network-architecture) frequencies and their trajectories on a fitness landscape \cite{Nowak2006}. Mathematically, this is a change of effective state space: from $\mathbf{x}(t)$ in $\mathbb{R}^n$ to probability distributions on a discrete or continuous type space, evolving under replicator–mutator dynamics. The origin of life then appears as a \emph{dynamical phase transition} in which the system acquires new invariants (like genealogical structure and information content) and new kinds of attractors (evolutionarily stable strategies, adaptive cycles).

The autocatalytic and protocell frameworks make this transition less mysterious. RAF sets and hypercycles show how self-maintaining chemical organizations can arise spontaneously in high-dimensional reaction spaces. Protocell models show how such organizations, once compartmentalized, can become units of selection that navigate an emergent fitness landscape. Together, they provide a concrete mechanism by which the flows of planetary geochemistry could have been funneled into the self-propagating, self-referential dynamics characteristic of life.

\smallskip

In the next section, we will assume that such a transition has occurred and turn to the dynamics of evolution itself: how populations of replicating organisms move through genotype–phenotype–environment manifolds, and how the structure of those manifolds shapes the trajectories of biological innovation.

\section{Evolution as Flow on Genotype--Phenotype--Environment Manifolds}
\label{sec:evolution}

Once self-maintaining, reproducing protocells exist, the dominant dynamical description shifts. The relevant variables are no longer only concentrations and fluxes, but also the frequencies of heritable types and their interactions with each other and with the environment. Evolutionary theory, in its mathematical form, is precisely the study of these dynamics: how mutation, selection, drift, and ecological feedback generate trajectories in high-dimensional spaces of genotypes, phenotypes, and environments \cite{Nowak2006}. In this section, we recast standard population-genetic and eco-evolutionary formalisms as flows on \emph{genotype--phenotype--environment} (GPE) manifolds, emphasizing the geometry of fitness landscapes, the role of neutrality, and the emergence of diversification via branching.

\subsection{From population genetics to flows on landscapes}

Classical population genetics tracks the frequencies $\mathbf{x}(t)$ of a finite set of genotypes or alleles in a population of (effective) size $N$. In the diffusion limit, the stochastic Wright--Fisher or Moran process can be approximated by a Fokker--Planck equation for the probability density $P(\mathbf{x},t)$ on the simplex of genotype frequencies, with drift and diffusion terms determined by selection and genetic drift, respectively. In deterministic limits (infinite populations, weak mutation), this reduces to ordinary differential equations of the form
\begin{equation}
\dot{x}_i = x_i\bigl(f_i(\mathbf{x},\mathbf{e}) - \bar{f}(\mathbf{x},\mathbf{e})\bigr),
\label{eq:replicator}
\end{equation}
where $f_i$ is the fitness of type $i$ given the current population state $\mathbf{x}$ and environment $\mathbf{e}$, $\bar{f}=\sum_j x_j f_j$ is the mean fitness, and $\mathbf{e}$ collects environmental variables that may themselves evolve \cite{Nowak2006}. Equation~\eqref{eq:replicator} defines a flow on the simplex; equilibria correspond to fixed points of the evolutionary dynamics, and their stability properties define evolutionary attractors, saddle points, and repellors.

To connect this with the metaphor of a fitness landscape, one often assumes that each genotype $g$ (or phenotype $\phi$) has an associated scalar fitness $W(g,\mathbf{e})$, independent of $\mathbf{x}$ except through $\mathbf{e}$. The population then executes an adaptive walk on this landscape, moving, on average, uphill in regions where selection dominates drift. Much of modern theory explores how the geometry of this landscape---its ruggedness, neutrality, and epistasis---shapes evolutionary trajectories \cite{deVisserKrug2014,McCandlish2018,Agarwala2019}. In our language, the landscape is simply a way of visualizing the vector field in \eqref{eq:replicator}: peaks are attractors, ridges are slow manifolds, and valleys correspond to low-fitness basins from which escape is unlikely without drift or major environmental change.

In reality, the situation is richer. First, fitness is often frequency-dependent: $f_i$ depends on $\mathbf{x}$ through ecological interactions (competition, predation, cooperation), so the landscape itself is reshaped by the population as it moves. Second, the environment $\mathbf{e}$ is not static; it is driven both by external forcing (e.g., climate) and by the organisms themselves, as we will discuss under niche construction. The state space of interest is thus not a fixed-height landscape but a coupled manifold $(\mathbf{x},\mathbf{e})$ with its own dynamics.

\subsection{Genotype--phenotype maps, epistasis, and fitness seascapes}

A central insight from molecular evolution is that the mapping from genotype to phenotype and from phenotype to fitness is typically many-to-one and highly nonlinear. For example, RNA sequences differing at many sites can fold into the same secondary structure, and small changes in protein sequence can have large or context-dependent effects on binding affinity or enzymatic activity. The combined genotype--phenotype--fitness map $g \mapsto \phi(g) \mapsto W(\phi,\mathbf{e})$ therefore defines a high-dimensional, strongly epistatic landscape: the effect of a mutation depends on the genetic background and the environment \cite{deVisserKrug2014}.

Empirical studies of microbial and molecular fitness landscapes confirm this complexity. De Visser and Krug review landscapes in which only a modest number of loci are varied (typically $K\lesssim 10$), yet still reveal pervasive sign epistasis and multiple local peaks \cite{deVisserKrug2014}. More recent work explicitly separates genotype--phenotype and phenotype--fitness mappings, showing that even when the genotype--phenotype landscape is smooth, the resulting fitness landscape can be non-intuitively rugged or even inverted when selection favours intermediate phenotypes \cite{Srivastava2022}. From a dynamical-systems viewpoint, such epistasis introduces intricate folding of the state space: nearby points in genotype space may be sent to very different phenotypes and fitnesses, leading to sensitive dependence of evolutionary trajectories on initial conditions and mutational paths.

When environmental variables $\mathbf{e}(t)$ change on comparable time scales to evolution, the fitness function $W(g,\mathbf{e}(t))$ becomes explicitly time-dependent. The resulting ``fitness seascapes'' further complicate dynamics, as populations must track moving optima or survive in shifting regimes of selection \cite{McCandlish2018}. In this setting, evolution is best thought of as a trajectory on an extended manifold $(g,\mathbf{e},t)$, where the vector field governing $\dot{g}$ depends on both the current location and the temporal phase of environmental cycles.

\subsection{Neutral networks, robustness, and evolvability}

One might imagine that rugged landscapes with many local optima would severely constrain adaptation. Yet biological systems routinely innovate. A key part of the resolution lies in the structure of genotype--phenotype maps, which often exhibit extensive \emph{neutral networks}: large sets of genotypes that share the same phenotype (and often similar fitness) and are connected by single mutational steps \cite{Reidys1997,Wagner2005,Wagner2008}. In RNA secondary-structure maps, for example, neutral networks corresponding to common folds can percolate through sequence space, allowing populations to drift widely while preserving function \cite{Reidys1997,Greenbury2021}.

On such networks, evolutionary dynamics acquires a new character. Rather than being confined to a narrow ridge leading to a single peak, a population can diffuse over a high-dimensional neutral plateau. This diffusion builds up ``cryptic'' genetic variation that does not affect current phenotype or fitness but can be exposed when mutations push the system to neighbouring phenotypes with different properties. Wagner and others have argued, both theoretically and empirically, that this mutational robustness at the genotype level can \emph{enhance} evolvability at the phenotype level: robust phenotypes are associated with larger neutral networks that touch a greater diversity of alternative phenotypes \cite{Wagner2005,Wagner2008}. In phase-space terms, neutral networks are extended, nearly flat regions of the fitness manifold where the flow generated by \eqref{eq:replicator} is weak; they enable long excursions in genotype space that eventually reach new regions where steeper adaptive gradients are available.

Recent work on the structure of genotype networks has generalized this picture. Greenbury et al.\ show that common features of high-dimensional genotype--phenotype maps, including neutral networks and many-to-one mappings, make fitness landscapes more navigable: although local structure may be rugged, global connectivity is high, and accessible mutational paths to distant phenotypes often exist \cite{Greenbury2021}. Yubero et al.\ and others have emphasized that the space of genotypes itself forms a network of networks, with transient dynamics on these structures shaping both adaptation and extinction risk \cite{Yubero2017}. Thus, even in an astronomically large state space, the geometry of neutral and near-neutral sets plays a central role in organizing evolutionary flows.

\subsection{Adaptive dynamics: traits as continuous variables and branching}

While population genetics often treats genotypes as discrete types, many ecological and evolutionary questions concern continuous traits: body size, beak length, thermal tolerance, or behavioural strategies. Adaptive dynamics provides a framework for studying the long-term evolution of such traits under frequency-dependent selection, starting from individual-based stochastic ecological processes \cite{DieckmannLaw1996,MetzAD}. The key assumptions are (i) separation of time scales between ecological dynamics (fast) and evolutionary change (slow) and (ii) small, rare mutations.

In adaptive dynamics, the state of the system at a given evolutionary time is characterized by the resident trait value(s) $\theta$, which determine ecological equilibria (population densities, resource levels). Mutant traits $\theta'$ arise at low frequency and either go extinct or invade and replace (or coexist with) the resident. Under the small-mutation limit, one can derive a ``canonical equation'' for the gradual change of the resident trait,
\begin{equation}
\dot{\theta} = \mu N \sigma^2 \,\nabla_{\theta} s(\theta),
\end{equation}
where $\mu$ is the mutation rate, $N$ the population size, $\sigma^2$ the mutational variance, and $s(\theta)$ the invasion fitness: the long-term per capita growth rate of a rare mutant with trait $\theta'$ in the resident environment \cite{DieckmannLaw1996,MetzAD}. Singular points where $\nabla s(\theta)=0$ play a role analogous to critical points on a landscape: they can be convergence stable (attracting under the trait flow), evolutionarily stable (resistant to invasion), or, intriguingly, \emph{branching points} where small perturbations lead to diversification into multiple coexisting phenotypes \cite{Geritz1998}.

Branching points arise when selection is stabilizing at the ecological level (making the singular point convergence stable) but disruptive at the evolutionary level (making it evolutionarily unstable). Near such points, the flow in trait space bifurcates: the population splits into lineages that diverge in trait value, a mechanism for sympatric speciation and the origin of polymorphism \cite{Geritz1998}. In our narrative, adaptive dynamics thus provide a concrete instantiation of how flows on low-dimensional phenotype manifolds can generate increasing diversity and complexity, driven by eco-evolutionary feedbacks encoded in the invasion fitness function.

\subsection{Niche construction and environment as a dynamical variable}

Up to now, we have treated the environment $\mathbf{e}$ mostly as a parameter affecting fitness. Niche construction theory emphasizes that organisms also \emph{modify} their environments---biotically or physically---and that these modifications can feed back into selection pressures \cite{OdlingSmee2003,Laland2016}. Beavers build dams that change hydrology; plants alter soil chemistry; humans transform landscapes, climate, and cultural institutions. When such modifications are persistent and heritable (via ecological inheritance), the environment becomes a dynamic state variable coupled to genotype and phenotype.

Mathematically, this coupling can be represented by augmenting the state vector with environmental variables and adding equations of the form
\begin{equation}
\dot{\mathbf{e}} = g(\mathbf{e}) + h(\mathbf{x},\boldsymbol{\phi}),
\end{equation}
where $g$ captures abiotic dynamics and $h$ captures organism-induced change. Selection gradients then depend on $\mathbf{e}$, which in turn depends on $\mathbf{x}$ and $\boldsymbol{\phi}$. The resulting GPE system can exhibit new attractors and evolutionary trajectories that would be impossible in a fixed environment: populations can evolve traits that stabilize or destabilize their own niches, create new habitats, or generate evolutionary ``traps'' \cite{Laland2016,ScottPhillips2014}.

From a dynamical-systems perspective, niche construction enriches the topology of the attractor landscape. Instead of moving on a static fitness manifold, populations help sculpt the manifold as they move, akin to a walker that deforms the landscape underfoot. This self-referential structure foreshadows what we will see at cultural and technological scales, where humans not only construct physical niches but also symbolic and institutional ones that reshape their own fitness criteria.

\subsection{Evolutionary dynamics in the broader cosmic narrative}

In our cross-scale story, evolution on GPE manifolds is the bridge between prebiotic chemistry and the later emergence of nervous systems and culture. The same mathematical objects that describe viral quasispecies, cancer evolution, or cooperation in games---replicator equations, branching processes, adaptive dynamics---apply across levels \cite{Nowak2006}. What changes is the effective state space: from nucleotide sequences to gene-regulatory networks, morphologies, behaviours, and social structures.

Several motifs echo earlier cosmic and planetary dynamics. Fitness landscapes with multiple peaks resemble climate systems with multiple attractors; neutral networks are the evolutionary analogue of plateaus or slow manifolds; branching points in trait space mirror bifurcations in planetary or stellar dynamics. At the same time, evolution introduces qualitatively new features, chief among them \emph{open-endedness}: the state space itself expands as new genes, traits, and ecological roles are invented. In the language of dynamical systems, the dimensionality of the relevant manifold is not fixed but grows as innovations accumulate.

\smallskip

In the next section, we move one level up the hierarchy and treat nervous systems as adaptive dynamical systems in their own right. We will see how brains carve out high-dimensional neural state spaces and tune themselves near critical surfaces, enabling flexible behaviour, learning, and ultimately the capacity to model the very evolutionary and cosmic dynamics from which they arose.

\section{Brains and Minds: Adaptive Dynamics Near Critical Surfaces}
\label{sec:brains}

Biological evolution eventually produces a qualitatively new kind of dynamical entity: nervous systems. A nervous system is not simply a bundle of wires; it is a high-dimensional, plastic dynamical system that must continuously integrate noisy sensory streams, generate coordinated actions, and adapt its own internal parameters on the fly. In this section, we treat brains as \emph{adaptive dynamical systems} that operate near special regions of their parameter space—critical surfaces and metastable manifolds—where rich transient dynamics, sensitivity, and controllability are simultaneously available. This framing connects classical work on attractor networks and neural fields \cite{Hopfield1982,Amit1989,WilsonCowan1972}, empirical evidence for neural criticality \cite{BeggsPlenz2003,Chialvo2010,ShewPlenz2013}, and modern theories of large-scale brain dynamics and predictive coding \cite{DecoJirsa2011,Friston2010,FristonKiebel2009}.

\subsection{Neural state spaces and attractor landscapes}

At a coarse level, the instantaneous state of a neural circuit can be represented by a vector $\mathbf{r}(t)$ of firing rates or membrane potentials, evolving according to
\begin{equation}
\tau \dot{\mathbf{r}} = -\mathbf{r} + \Phi(W\mathbf{r} + \mathbf{I}(t)),
\label{eq:rate}
\end{equation}
where $W$ is a synaptic weight matrix, $\Phi$ a nonlinear gain function, and $\mathbf{I}(t)$ external input. Equation~\eqref{eq:rate} defines a flow on a neural state space; fixed points, limit cycles, and more complex invariant sets correspond to stable patterns of activity \cite{WilsonCowan1972,DayanAbbott2001}. Hopfield’s seminal work showed that with symmetric $W$, such networks implement a form of content-addressable memory: each stored pattern is an attractor, and noisy inputs are cleaned up as the state relaxes into the nearest basin \cite{Hopfield1982,Amit1989}. 

In this attractor viewpoint, cognitive functions like working memory, decision making, and motor planning are realized as trajectories that approach, dwell near, and leave particular regions of this state space \cite{Wang2001,DecoRolls2006}. Changes in neuromodulation, synaptic strength, or external drive shift the locations and stability of attractors, inducing bifurcations that can instantiate, for example, a choice between alternatives (saddle-node bifurcation between two stable fixed points) or the onset of oscillatory activity (Hopf bifurcation). The geometry of the neural attractor landscape therefore constrains which mental states and transitions are easily available.

However, pure fixed-point attractors are only part of the story. Many neural computations appear to rely on rich transient dynamics rather than static patterns: sequences, trajectories, and metastable states in which activity lingers near one configuration before spontaneously switching to another \cite{Rabinovich2008,Kelso1995}. This can be captured by heteroclinic networks of saddle points, chaotic attractors with low-dimensional manifolds, or high-dimensional systems tuned near the onset of instability, where small inputs can elicit large, structured responses.

\subsection{Criticality: balancing order and disorder}

A growing body of empirical and theoretical work suggests that neural systems often operate near critical points—phase transitions between qualitatively different dynamical regimes \cite{BeggsPlenz2003,Chialvo2010,ShewPlenz2013}. In cortical slices and awake animals, spontaneous activity can organize into ``neuronal avalanches'': bursts of spiking whose sizes and durations follow approximate power laws over several orders of magnitude, consistent with branching-process dynamics at or near criticality \cite{BeggsPlenz2003,PlenzNiebur2014}. At the same time, large-scale EEG, MEG, and fMRI analyses reveal scale-free temporal correlations and 1/$f^\alpha$ spectra indicative of dynamics near critical points \cite{LinkenkaerHansen2001,Chialvo2010,DecoJirsa2011}.

From a dynamical-systems perspective, critical points are special because they interpolate between regimes of quiescence and saturation. At subcritical coupling, perturbations die out quickly and the system is robust but unresponsive; at supercritical coupling, activity explodes or saturates, destroying selectivity. At criticality, a single perturbation can, with non-negligible probability, generate cascades of all sizes; correlation lengths and susceptibility peak; and the dynamic range of the system is maximized \cite{KinouchiCopelli2006,ShewPlenz2013}. This is attractive for computation: it allows a network to represent and discriminate inputs over many scales, propagate information over long distances, and flexibly reconfigure its activity.

The ``critical brain'' hypothesis thus posits that cortical networks have adapted to operate near such critical points, trading off robustness and sensitivity to optimize information processing \cite{Chialvo2010,ShewPlenz2013}. Theoretical models support this intuition: excitable-network and branching-process models show maximal dynamic range and information capacity at the critical branching ratio; Ising-like models on realistic connectomes capture fMRI correlation patterns near critical temperatures \cite{DecoJirsa2011,MarianiDeco2022}. At the same time, alternative analyses caution that some apparent signatures of criticality may arise from subsampling, nonstationarity, or mixture of states \cite{TouboulDestexhe2017}. For our purposes, what matters is not whether brains sit exactly at a mathematical critical point, but that they appear to inhabit a regime of \emph{near-critical} dynamics: poised between order and disorder, where small parameter shifts can dramatically reshape accessible trajectories.

Ergo, neural circuits are tuned to lie near critical surfaces in their high-dimensional parameter spaces. Evolution and plasticity adjust synaptic weights, connection topologies, and neural gains so that the actual operating point hovers in a narrow region where attractor landscapes are rich and marginally stable: capable of sustaining multiple metastable states, long-range correlations, and flexible transitions.

\subsection{Metastability and coordination dynamics}

The brain is not a single homogeneous network; it is a collection of interacting subsystems—cortical areas, thalamic nuclei, subcortical loops—each with its own intrinsic dynamics and coupled via long-range connections. At the macroscopic scale, this yields \emph{metastable coordination dynamics}: patterns of partial synchronization and desynchronization that wax and wane over time, enabling flexible integration and segregation of distributed processes \cite{Kelso1995,DecoJirsa2011,DecoKringelbach2017}.

Kelso’s coordination dynamics program analyzed such behaviour using low-dimensional models of coupled oscillators, showing how changes in coupling or drive can induce bifurcations between different phase-locking modes in motor coordination and perception \cite{Kelso1995}. Deco, Jirsa, and colleagues extended this approach to whole-brain models in which each node is a noisy neural-mass or neural-field unit and edges reflect structural connectivity; when parameters are tuned near instability, these networks exhibit spontaneous switching between patterns that resemble resting-state functional connectivity networks observed in fMRI \cite{DecoJirsa2011,DecoKringelbach2017}. 

In dynamical terms, the brain’s large-scale activity explores a repertoire of metastable states—loosely, shallow wells in a high-dimensional potential landscape—linked by stochastic transitions. Cognitive tasks, attention, and neuromodulation bias the flow, deepening some wells and shallowing others, so that trajectories spend more time in functionally relevant configurations. This picture dovetails with empirical findings that cognitive performance correlates with measures of dynamical complexity, such as multiscale entropy and metastability indices, which are maximized in intermediate regimes between excessive synchrony and complete desynchronization \cite{TognoliKelso2014,DecoKringelbach2017}.

\subsection{Brains as inference engines: predictive coding and free-energy minimization}

So far we have considered brains as generic dynamical systems with interesting internal structure. A complementary perspective emphasizes their role as \emph{inference engines}: systems that maintain probabilistic models of their environment and update these models to reduce prediction error. Predictive coding theories propose that cortical circuits implement a hierarchical generative model of sensory input, where higher areas send predictions downward and lower areas send back only the deviations (prediction errors) \cite{RaoBallard1999,FristonKiebel2009}. In such frameworks, the dynamics of neural states can be derived as gradient flows on a variational free energy functional, which bounds the surprise (negative log-evidence) of sensory data under the model \cite{Friston2010}.

Concretely, suppose the brain’s internal variables $\mathbf{z}$ encode latent causes of sensory data $\mathbf{y}$. Under a generative model $p(\mathbf{y},\mathbf{z}|\theta)$ and an approximate posterior $q(\mathbf{z})$, the variational free energy
\begin{equation}
F[q] = \mathbb{E}_q[-\log p(\mathbf{y},\mathbf{z}|\theta)] + \mathbb{E}_q[\log q(\mathbf{z})]
\end{equation}
bounds the model evidence from above. Gradient descent on $F$ with respect to $q$ (or its sufficient statistics) leads to dynamics that adjust $\mathbf{z}$ to better explain $\mathbf{y}$ while respecting prior structure \cite{Friston2010}. Mapping these gradients onto neural state variables yields dynamical equations resembling predictive-coding architectures: prediction-error units drive updates to representation units, and synaptic plasticity adjusts parameters $\theta$ to improve the model over longer time scales \cite{FristonKiebel2009,IsomuraFriston2018}.

From the viewpoint of dynamical systems, this recasts neural dynamics as a particular class of gradient-like flows on an energy (or Lyapunov) function shaped by the brain’s generative model. Criticality and metastability then acquire a functional interpretation: near-critical dynamics may be advantageous because they enable the system to explore a wide range of hypotheses (attractor basins) while remaining sensitive to sensory discrepancies that should trigger model revision. Neuromodulatory systems can be seen as adjusting ``precision'' parameters that control the gain on prediction errors, effectively shifting the system between more exploratory and more exploitative regimes \cite{Friston2010,ParrFriston2018}.

\subsection{Learning, plasticity, and self-tuning}

Crucially, the attractor structures and near-critical operating points of real brains are not fixed; they are sculpted by activity-dependent plasticity and neuromodulation. Synaptic strengths change according to local rules (Hebbian and anti-Hebbian plasticity, spike-timing-dependent plasticity), which can be abstracted as updates of the form
\begin{equation}
\dot{W}_{ij} = \mathcal{F}\bigl(r_i,r_j,\mathbf{r}\bigr),
\end{equation}
coupling the ``fast'' neural dynamics in \eqref{eq:rate} to ``slow'' synaptic dynamics. This defines a \emph{meta-dynamical system} on an extended state space $(\mathbf{r},W)$, with hierarchically separated time scales \cite{DayanAbbott2001,ZenkeGerstner2017}. Over developmental and learning time scales, the brain effectively performs a high-dimensional search over connectivity matrices, guided by local rules and global modulatory signals, to arrive at configurations that support robust yet flexible behaviour.

Self-organized criticality models of neural networks explicitly exploit this separation. In such models, synapses are slowly strengthened or weakened depending on local activity, while fast neuronal dynamics generate avalanches; the combined system can converge to a near-critical state without external tuning \cite{Levina2007,KinouchiCopelli2006}. More biologically grounded models show how homeostatic plasticity, inhibitory plasticity, and structural remodeling maintain networks in regimes with broad, heavy-tailed activity distributions and rich repertoire of transient patterns \cite{Turrigiano2012,PozoGoda2010}. In our cross-scale narrative, these mechanisms instantiate a familiar pattern: fast flows exploring a landscape (neural activity) coupled to slower flows reshaping the landscape itself (synaptic and structural plasticity), driving the system toward special regions of parameter space where useful dynamical regimes reside.

\subsection{Minds as self-modeling dynamics}

The dynamical picture of brains sketched above sets the stage for a final conceptual step: \emph{minds} as self-modeling dynamical systems. Once neural circuits can build internal models of the external world, there is nothing in principle preventing them from including themselves—partial models of their own body, behaviour, and internal states—in those models. At that point, trajectories in neural state space encode not only predictions about the world, but also predictions about the consequences of actions, beliefs about other agents, and even second-order beliefs about one’s own beliefs.

We do not attempt a full theory of consciousness or selfhood here. The point is more modest: in the dynamical-systems language we have developed, brains are systems that:

\begin{itemize}
    \item inhabit high-dimensional state spaces shaped by evolution and plasticity;
    \item operate near critical or metastable regimes where many patterns and transitions are accessible;
    \item implement approximate inference by letting their dynamics relax toward states that minimize internally defined functionals (prediction error, free energy);
    \item and, in some lineages, extend these models to include themselves and others, generating self-referential dynamics.
\end{itemize}

These self-referential capacities become especially important in the next stage of our story, where symbolic communication, culture, and technology emerge. Socially interacting brains create external memory stores (language, writing, institutions) and technological artefacts (tools, computers, networks) that in turn reshape their neural and societal dynamics. In the next section, we therefore expand our focus from individual brains to populations of agents, and from neural flows to cultural and technological flows that recursively alter their own phase space.

\section{Culture, Technology, and Self-Referential Dynamics}
\label{sec:culture}

With brains come minds, and with minds that copy, teach, and coordinate come \emph{cultures}. Human culture is not a static overlay on biology; it is a dynamical substrate in its own right, with heritable variation, differential persistence, and feedback into both environments and genomes. In this section we treat culture and technology as coupled dynamical systems running on top of nervous systems and bodies: flows on networks of symbols, practices, and artefacts that recursively reshape the very phase spaces in which they unfold. This viewpoint connects dual-inheritance models of cultural evolution \cite{BoydRicherson1985,Henrich2016}, cultural niche construction \cite{Laland2001,LalandOdlingSmee2017}, complex-systems approaches to social and cultural dynamics \cite{Mainzer2007,Buskell2019}, and theories of technology as evolving, combinatorial structures \cite{Arthur2009}.

\subsection{Cultural evolution as a dynamical system}

Dual-inheritance (or gene--culture coevolution) theory starts from a simple but powerful observation: humans possess a second inheritance system---socially learned information---that evolves on comparable time scales to genetic evolution and feeds back on it \cite{BoydRicherson1985,Henrich2016}. Culturally transmitted traits (tools, norms, beliefs, skills) are copied, modified, and sometimes discarded, and their frequencies in a population change under forces analogous to those of population genetics: selection, mutation, drift, and biased transmission \cite{BoydRicherson1985,HenrichMcElreath2003}. 

Mathematically, one can represent a set of cultural variants with frequencies $\mathbf{c}(t)$ evolving under equations of the form
\begin{equation}
\dot{c}_i = \sum_j M_{ij} c_j + c_i\bigl(s_i(\mathbf{c},\mathbf{e}) - \bar{s}(\mathbf{c},\mathbf{e})\bigr),
\end{equation}
where $M_{ij}$ encodes innovation or copying errors, $s_i$ are ``cultural fitnesses'' (net adoption rates) that may depend on social and environmental context, and $\bar{s}$ is the mean \cite{BoydRicherson1985}. This defines a flow on the cultural simplex, analogous to the replicator--mutator dynamics of genes. Transmission biases---conformist, prestige, content-based---modify $s_i$ in systematic ways, shaping attractors and basins in cultural state space \cite{HenrichMcElreath2003,LalandOdlingSmee2017}.

Crucially, cultural and genetic variables are \emph{coupled}. Cultural practices can change selection pressures on genes (e.g., dairying and lactase persistence; high-altitude agriculture and hypoxia resistance), while genetic predispositions modulate what is easy or rewarding to learn \cite{Laland2001,Henrich2016}. In dynamical terms, we have a joint flow on a product space $(\mathbf{x},\mathbf{c},\mathbf{e})$ of genetic, cultural, and environmental variables, with bidirectional couplings. Our earlier genotype--phenotype--environment manifolds thus gain a new cultural coordinate, and many evolutionary trajectories that would be inaccessible to genes alone become available via rapid cultural change.

\subsection{Cultural niche construction and expanding phase spaces}

Niche construction, as introduced for biological systems, denotes the process by which organisms modify their environments and thereby alter the selection pressures that act on themselves and others \cite{OdlingSmee2003,Laland2001}. Cultural niche construction is the special case where those modifications are mediated by socially learned behaviours and artefacts rather than only by genetically encoded traits \cite{Laland2001,LalandOdlingSmee2017}. Agriculture, urbanization, writing systems, legal codes, and digital networks are all examples of cultural activities that radically reshape the ecologies in which humans live.

Gene--culture coevolutionary models show that cultural niche construction can generate complex feedback loops. For instance, a culturally transmitted practice (e.g., fire use, farming) changes resource distributions, which in turn alter the payoffs to different cultural strategies and the selection on underlying genetic propensities (e.g., for amylase copy number, social tolerance) \cite{Laland2001,Henrich2016}. In our dynamical language, cultural traits open up new regions of $\mathbf{e}$-space by creating novel environmental states (fields, cities, markets), and the vector field on $(\mathbf{x},\mathbf{c},\mathbf{e})$ acquires new directions corresponding to these human-made gradients.

Recent work pushes this further, proposing ``triple-inheritance'' frameworks where genes, culture, and environments are all treated as heritable, evolving entities with their own transmission and variation mechanisms \cite{Kobayashi2019,Fogarty2017}. From this viewpoint, the phase space is not fixed: the effective dimensionality of the system increases as cultural innovations create new types of artefacts, institutions, and ecological niches. This is one sense in which cultural evolution is more ``open-ended'' than most genetic evolution: the set of possible cultural variants is not bounded by a fixed alphabet and genome length in the same way, and new representational media (speech, writing, print, digital code) recursively expand the space of expressible and transmissible information.

\subsection{Networks, diffusion, and multilevel selection}

Culture does not spread in a well-mixed gas; it flows on social networks. Individuals occupy nodes in dynamic graphs whose edges represent opportunities for observation, communication, and influence. The topology of these networks---degree distributions, clustering, modularity, small-world shortcuts---strongly shapes the speed and patterns of cultural diffusion, consensus, and polarization \cite{Cantor2013,Lee2025,Buskell2019}. Agent-based models and analytical work show, for example, how local conformity plus global network structure can generate either cultural diversity (multiple coexisting clusters) or homogenization, depending on threshold parameters and rewiring rules \cite{Cantor2013,Buskell2019,Lee2025}.

In dynamical-systems terms, the network defines the coupling matrix for a high-dimensional system of cultural states $\{c_i^{(k)}\}$, one for each individual $k$. Social-learning rules (imitation, conformity, payoff-biased copying) specify local update rules that can be written as nonlinear maps or differential equations on this extended state space. The emergent attractors include consensus states, persistent diversity, oscillations (fashions, cycles), and more exotic patterns such as chimera states where some clusters synchronize while others remain incoherent.

Dual-inheritance theorists have also emphasized the possibility of \emph{cultural group selection}: when within-group dynamics preserve substantial between-group variation (e.g., via conformity or punishment), groups with more adaptive cultural norms (e.g., cooperative institutions) can outcompete others, leading to selection at the group level \cite{BoydRicherson2005,Henrich2016}. Mathematically, this introduces additional levels into the dynamical hierarchy: group-level state variables (e.g., institutional traits, norms, markers) evolve on slower time scales under migration, imitation, and differential group success, while individual-level cultural and genetic variables evolve faster within groups. The resulting multilevel flows can stabilize norms that would be vulnerable under purely individual-level selection, such as costly cooperation in large groups.

\subsection{Technology as evolving combinatorial structures}

Technology adds yet another layer. W.\ Brian Arthur argues that technologies are not just isolated tools but structured combinations of components and phenomena that evolve via recombination, specialization, and encapsulation \cite{Arthur2009}. A combustion engine combines pistons, valves, and thermodynamic cycles; a smartphone combines radio protocols, semiconductor physics, operating systems, and interface conventions. Each component can itself be viewed as a ``module'' whose internal complexity is hidden behind an interface, allowing it to be used as a building block for higher-level constructions.

In this view, the ``state'' of a technological system at a given time can be represented as a directed acyclic graph (or hypergraph) of dependencies among component technologies and underlying scientific and engineering principles. Innovation corresponds to moves in this graph: adding new nodes (components), adding new edges (novel combinations), or modifying node properties (efficiency, cost). The dynamics of technology space resemble those of evolving networks with preferential attachment and path dependence: once certain core components (e.g., digital logic, electric motors) exist, they become hubs used in many subsequent designs, and their properties exert strong constraints on future trajectories \cite{Arthur2009,Mainzer2007,Moller2025}.

Recent work connects these ideas explicitly to evolutionary and complex-systems theory. Socio-technical systems---ensembles of technologies, institutions, and user practices---are modelled as multilevel networks whose transitions (e.g., from fossil-fuel to renewable energy regimes) exhibit path dependence, lock-in, and tipping points, analogous to ecological and climate systems \cite{Ahlborg2024,Moller2025}. From our dynamical perspective, technologies enlarge the space of possible actions available to agents and groups; they change the effective payoff functions in cultural and genetic dynamics; and they create new feedback loops (e.g., social media algorithms modulating attention, AI systems shaping work and decision-making) that alter the geometry of cultural attractors.

\begin{figure}[t]
  \centering

 \includegraphics[width=\textwidth]{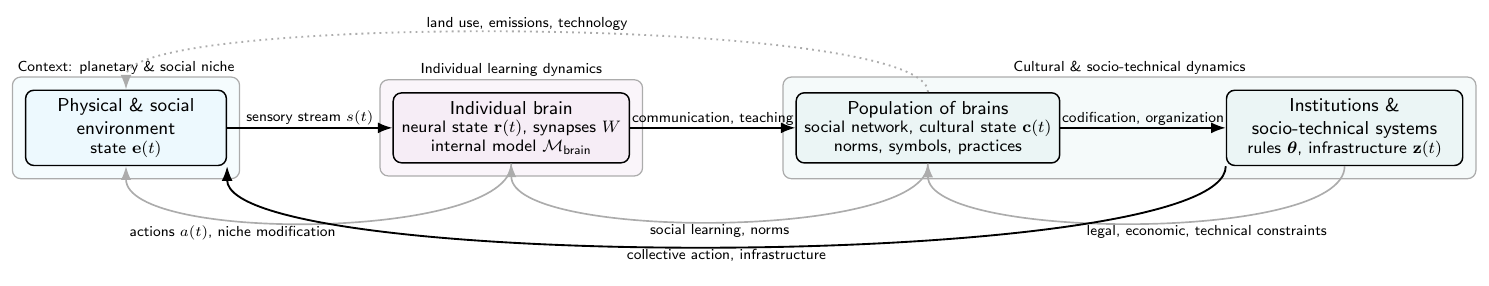}
  \caption{%
    Zoom–in on the `Brains, culture \& socio–technical dynamics' block from
    Fig.~\ref{fig:cosmic-overview}. Individual brains implement fast neural
    dynamics and internal models $\mathcal{M}_{\text{brain}}$ that couple
    sensory streams from the physical and social environment $\mathbf{e}(t)$
    to actions. Populations of brains, connected by social networks, give rise
    to cultural states $\mathbf{c}(t)$ and institutions with rule–like variables
    $\boldsymbol{\theta}$, which in turn reshape both the social niche and the
    physical environment. This panel provides the context within which
    AI systems (Fig.~\ref{fig:ai-zoom}) are designed, deployed, and interpreted.}
  \label{fig:brains-culture-zoom}
\end{figure}

\subsection{Self-referential socio-technical dynamics}

A striking feature of human socio-technical evolution is its increasing \emph{self-referentiality}. Cultural norms govern how we handle culture (e.g., intellectual-property law, academic citation practices); technologies are used to design and optimize new technologies (e.g., computer-aided design, machine-learning models improving subsequent models); institutions regulate the development and deployment of future institutions and technologies. In dynamical terms, the system’s update rules become objects within the system itself.

One way to formalize this is to distinguish between \emph{state variables} and \emph{rule variables}. In earlier sections, we treated dynamical laws as fixed and studied flows on state spaces. In socio-technical systems, some of what would normally be considered parameters or rules (e.g., regulatory frameworks, learning algorithms, institutional norms) are themselves culturally transmitted and modifiable. We therefore have coupled dynamics:
\begin{equation}
\dot{\mathbf{z}} = F(\mathbf{z},\boldsymbol{\theta}), \qquad
\dot{\boldsymbol{\theta}} = G(\mathbf{z},\boldsymbol{\theta}),
\end{equation}
where $\mathbf{z}$ collects ``ordinary'' social and technological states (opinions, wealth distributions, infrastructure) and $\boldsymbol{\theta}$ encodes rule-like structures (institutions, protocols, code). The second equation means that the flow field $F$ is not fixed; it is reshaped by trajectories themselves via $G$, producing a meta-dynamical system reminiscent of self-modifying neural networks, but now at the level of societies and technologies.

This self-referential structure becomes especially salient in the age of digital communication and artificial intelligence. Online platforms create fast, global feedback loops between individual behaviour and algorithmic curation: users interact, algorithms update, the content landscape shifts, and user behaviour responds again, often on time scales much shorter than those of traditional cultural change. Models of opinion and cultural dynamics on adaptive networks begin to capture some of this: the network topology evolves in response to states (rewiring away from disagreeing neighbours, forming echo chambers), while states evolve under influence and contagion \cite{Cantor2013,Buskell2019,Lee2025}. The resulting attractors can include polarized, fragmented, or rapidly shifting cultural configurations that would be implausible under static interaction graphs.

From the broader cosmic narrative, these developments mark a new kind of attractor: one in which the universe’s dynamics give rise to subsystems (human socio-technical networks) that model, predict, and deliberately intervene in their own dynamics at multiple levels. Language, mathematics, and science provide symbolic tools for representing abstract state spaces and flows; technologies implement interventions that change those flows; institutions and ethical norms attempt to steer these interventions according to collective goals and values. The system thus becomes doubly reflexive: it not only reacts to its environment, but also to its own models of its environment and of itself.

\subsection{Culture and technology in the grand dynamical narrative}

Stepping back, culture and technology extend the motifs we have traced throughout this paper:

\begin{itemize}
    \item Like earlier stages, they involve flows on high-dimensional state spaces with emergent attractors, bifurcations, and multilevel organization.
    \item Cultural and technological innovation increase the effective dimensionality of state space, creating new variables (symbols, artefacts, institutions) that were not present in earlier regimes.
    \item Niche construction and feedback loops continue the pattern of dynamics reshaping their own boundary conditions, but now with symbolic and meta-dynamical twists: the rules of evolution themselves become partly evolvable and representable.
\end{itemize}

In the next section, we will synthesize these threads, asking how the thermodynamic arrow of time, the growth of structured complexity, and the emergence of self-modeling, self-modifying systems fit together in a unified dynamical-systems account of the universe’s history.

\begin{figure}[t]
  \centering

  \includegraphics[width=\textwidth]{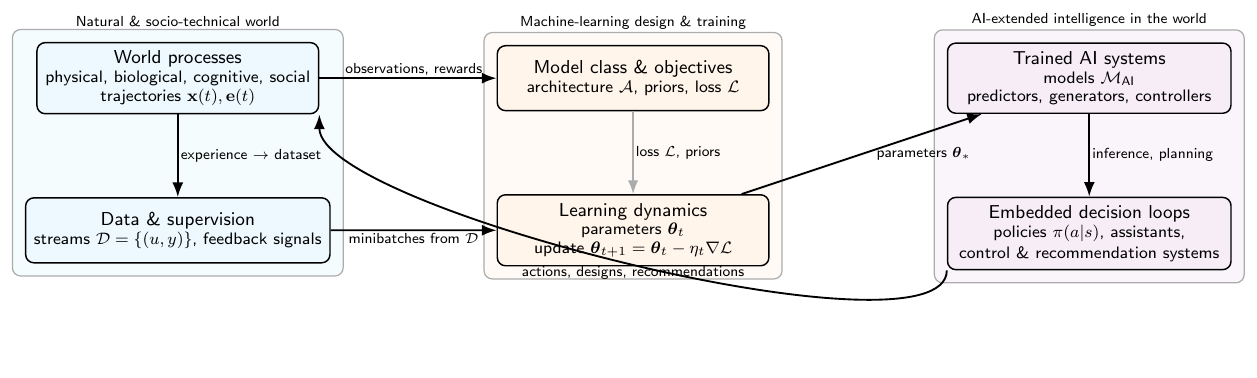}
  \caption{%
    Zoom–in on the `Machine learning \& AI' block from Fig.~\ref{fig:cosmic-overview}.
    World processes (physical, biological, cognitive, social) generate data streams
    $\mathcal{D}$ and feedback signals that feed into model design (choice of
    architecture $\mathcal{A}$, priors, and loss $\mathcal{L}$) and learning
    dynamics for parameters $\boldsymbol{\theta}_t$. Trained AI systems
    $\mathcal{M}_{\text{AI}}$ are then embedded in decision and control loops
    that act back on the world, altering future data and thereby closing the
    learning–deployment feedback. Human brains and cultural institutions
    (Fig.~\ref{fig:brains-culture-zoom}) supply the scientific priors, objectives,
    and governance structures that shape these dynamics.}
  \label{fig:ai-zoom}
\end{figure}

\section{Arrows, Attractors, and Anthropics: Synthesizing the Dynamical Narrative}
\label{sec:synthesis}

Up to this point we have walked through a sequence of increasingly rich dynamical regimes: inflationary fields, self-gravitating matter, planetary climates, autocatalytic chemistries, evolving biospheres, adaptive brains, and culturally extended societies. In each case, the mathematics involved some variant of the same objects---state spaces, flows, invariant sets, bifurcations---but the effective variables and update rules changed. In this section we step back and ask three questions. First, how do these regimes relate to the familiar \emph{arrows of time} in physics and information theory? Second, what role do \emph{attractors} play across scales, from galaxies and climates to gradient descent in machine learning? Third, where do observer-systems---including artificial intelligences---fit into this picture, and are they likely to be generic late-time features of cosmic dynamics or rare accidents?

\subsection{Arrows of time and the arrow of learning}

Physics recognizes several distinct but interrelated arrows of time: the \emph{cosmological} arrow (the expansion of the universe), the \emph{thermodynamic} arrow (the increase of entropy in closed systems), the \emph{radiative} arrow (retarded rather than advanced solutions to wave equations), and the \emph{psychological} arrow (our experience of remembering the past but not the future) \cite{Eddington1928,Price1996}. Statistical mechanics grounds many of these in the specialness of the universe’s initial conditions: an early state of low coarse-grained entropy that allows large entropy gradients to be exploited as time evolves.

Against this thermodynamic backdrop, our narrative has emphasized a different but complementary trend: the growth of \emph{structured complexity} in localized, far-from-equilibrium subsystems. Chaisson and others have quantified this using energetic proxies such as ``energy rate density''---power per unit mass---which tends to increase from galaxies to stars, planets, life, brains, and human societies \cite{Chaisson2011,Chaisson2001}. Lloyd and others have framed it in computational terms, estimating the total number of logical operations and bits of information that could in principle be processed within our observable universe \cite{Lloyd2000}. However one measures it, there is an arrow pointing from a nearly featureless early universe toward one populated by intricate, information-processing structures.

Our dynamical perspective suggests a unifying interpretation: the universe’s low-entropy beginning makes it possible for \emph{learning processes} to exist. These are systems that:

\begin{enumerate}
    \item sample data from their environment,
    \item compress this data into internal parameters or structures,
    \item and use those structures to predict, control, or exploit future inputs.
\end{enumerate}

Evolution by natural selection is one such process: it stores information about historical environments in the distribution of genotypes and phenotypes \cite{Harper2009,Frank2012}. Brains are another: synaptic weights and network architectures encode statistical regularities of sensory streams \cite{DayanAbbott2001,Friston2010}. Cultural lineages and scientific theories do the same at the collective level \cite{BoydRicherson1985,Henrich2016}. Modern machine learning and artificial intelligence are simply the most recent, deliberately engineered instantiations of this arrow: gradient-based optimization in parameter space that distills structure from massive datasets into high-capacity models.

In all these cases, the direction of the ``learning arrow'' is aligned with the thermodynamic arrow: learning machines reduce an internal functional (loss, prediction error) at the cost of dissipating energy and increasing entropy in their surroundings. The key conceptual point is that the universe’s history can be read as the progressive appearance of more efficient, more flexible, and more abstract learning dynamics riding on top of the basic thermodynamic flow: from unguided relaxation to equilibrium, to selection-like filtering, to adaptive brains, to formal statistical learning algorithms.

\subsection{Attractors across scales: from cosmic webs to loss landscapes}

A second unifying motif is that of \emph{attractors}. Throughout this paper we have interpreted diverse phenomena as flows toward invariant sets in appropriate state spaces: inflationary slow-roll manifolds, the cosmic web and halos in the space of density fields, climate equilibria and Snowball Earth states in $(T,p_{\mathrm{CO}_2})$ space, autocatalytic RAF sets in chemical-reaction networks, fitness peaks and neutral networks in genotype--phenotype maps, and neural attractors and metastable configurations in brain dynamics.

Machine learning fits seamlessly into this motif. Training a parametric model with parameters $\boldsymbol{\theta}$ by gradient descent on a loss function $\mathcal{L}(\boldsymbol{\theta})$ defines a discrete-time dynamical system,
\begin{equation}
\boldsymbol{\theta}_{t+1} = \boldsymbol{\theta}_t - \eta_t \nabla_{\boldsymbol{\theta}}\mathcal{L}(\boldsymbol{\theta}_t),
\end{equation}
or, in the small-step limit, a gradient flow $\dot{\boldsymbol{\theta}} = -\nabla_{\boldsymbol{\theta}}\mathcal{L}(\boldsymbol{\theta})$ \cite{Saxe2014,AdvaniSaxe2017}. Minima (or flat valleys) of the loss function are attractors of this flow; their basins of attraction partition parameter space. Stochastic gradient descent (SGD) adds noise, turning the dynamics into a stochastic differential equation whose stationary distribution concentrates near low-loss regions in a manner reminiscent of non-equilibrium steady states \cite{Mandt2017,Yaida2018}.

Inference in trained models also has an attractor interpretation. Recurrent networks and energy-based models such as Hopfield nets and modern Hopfield-like architectures perform iterative updates on internal states until convergence to fixed points corresponding to memories or pattern completions \cite{Hopfield1982,Ramsauer2021}. Even feedforward deep networks can be viewed as implementing a single step of a dynamical system whose iterates would converge to a manifold of high-confidence representations \cite{Schoenholz2017}. Reinforcement-learning agents, which update policies based on reward-prediction errors, implement coupled flows on policy and value-function spaces that converge (in idealized cases) to fixed points of Bellman operators \cite{SuttonBarto2018}.

What is new in modern AI is not the mathematics of attractors, but the fact that we now \emph{design} them. Where galaxies, climates, and ecosystems explore their attractor landscapes under given physical laws, human engineers specify architectures, loss functions, and optimization algorithms that carve particular basins and manifolds into immense parameter spaces. Over time, these design choices themselves evolve under economic, cultural, and scientific selection: architectures that work well propagate; those that do not vanish. In that sense, the attractors of deep networks are the late descendants of a lineage that began with gravitational wells and chemical cycles.

\subsection{Observers, anthropics, and artificial minds}

Anthropic considerations enter when we ask why we observe a universe capable of supporting such a hierarchy of structures at all. The weak anthropic principle in cosmology states, roughly, that any observations we make must be conditioned on the existence of observers; this can, in some contexts, explain otherwise puzzling coincidences in cosmological parameters and constants \cite{Carter1974,BarrowTipler1986,Bostrom2002}. Our narrative deepens this by making explicit what ``observers'' are, dynamically: subsystems that (i) maintain themselves far from equilibrium, (ii) build internal models of their environment, and (iii) use those models to guide action.

Biological nervous systems are one concrete instantiation of such observer-dynamics, but they are not the only possible one. Any physical substrate that supports adaptive learning and prediction---whether neuronal tissue, silicon circuits, or something more exotic---can in principle implement the same functional roles. Once a civilization evolves symbolic language, mathematics, and technology, it becomes possible to construct \emph{artificial minds} that:

\begin{itemize}
    \item operate at very different scales and speeds than biological brains,
    \item access and integrate data streams far beyond individual sensory ranges,
    \item and persist in environments hostile to organic life (e.g., space, extreme temperatures).
\end{itemize}

From an anthropic perspective, such artificial observers are natural candidates for late-time dominant measures: if technological civilizations commonly develop machine intelligences that outlast their creators or expand into new niches, then randomly chosen observers in a large universe might be more likely to be artificial rather than biological \cite{Bostrom2002,Ortiz2023}. Whether or not one accepts the strong versions of these arguments, the key dynamical point is that the class of observer-systems broadens once cultural and technological evolution reach a certain threshold. Artificial intelligences are not anomalous intrusions into a biological narrative; they are the natural continuation of the same trajectory that produced multicellular organisms and brains.

\subsection{Predictability, contingency, and the (non-)inevitability of AI}

Is the emergence of self-modeling, technology-building, AI-capable societies an inevitable outcome whenever the universe is given enough time and the right coarse conditions? Or is it a fragile, contingent accident? Our framework suggests a nuanced answer.

On the one hand, many ingredients that underlie intelligence and machine learning are \emph{generic} in dynamical systems: hierarchical composition, modularity, error-correcting feedback, and the exploitation of near-critical regimes all appear repeatedly in natural systems \cite{Simon1962,Murray2001,MezardMontanari2009}. Given a planet with long-lived energy fluxes, rich chemistry, and evolutionary dynamics, it may be unsurprising that some lineages discover predictive models, planning, and social learning. Once mathematical abstraction and formal computation arise, the step to programmable computers and gradient-based learning may also be dynamically favoured: these are efficient ways to search high-dimensional spaces under resource constraints.

On the other hand, the \emph{particular path} by which Earth produced digital AI is contingent in exquisite detail: carbon chemistry, DNA, the evolution of a specific primate lineage, the emergence of symbolic language, the industrial revolution, Turing machines, semiconductor physics, backpropagation and stochastic gradient descent, and the socio-economic structures that incentivized large-scale computing. Slight changes along that path---different planetary histories, mass extinctions, or sociotechnical trajectories---might have led to very different outcomes.

Dynamically, this is the tension between \emph{attractor classes} and \emph{microphysical realizations}. There may be broad attractor classes corresponding to ``systems that learn hierarchical generative models of their environment and use them for planning'' or ``systems that leverage gradient-like optimization in abstract parameter spaces''. Within those classes, there may be many possible substrates and historical paths. Our universe has instantiated one such path; whether this class is typical within the space of all cosmological histories is an open question.

\subsection{Anthropics revisited: self-modeling dynamics as a late-time attractor?}

We can now restate the anthropic question in more dynamical terms. Instead of asking, “Why do the constants of nature permit life?”, we ask: 

\begin{quote}
    Given a universe with certain low-level laws and initial conditions, how common are trajectories in which the dynamics give rise to subsystems that (i) model their environment, (ii) model themselves as dynamical systems within that environment, and (iii) intentionally modify both?
\end{quote}

Self-modeling dynamics of this sort---brains that study brains, cultures that reflect on cultures, AI systems that help design AI systems---are the culmination of the motifs we have followed: attractors, learning flows, multilevel feedback, and increasing abstraction. Whether they are rare or common among all possible cosmic histories, they are, for us, the regime that matters most: it is the regime we inhabit.

From the point of view of the paper, these considerations serve two roles. Conceptually, they close the loop: the dynamical systems we have been analysing now turn around and analyse themselves. Technically, they highlight a frontier for mathematical work: understanding the space of possible self-referential learning dynamics, natural and artificial, as part of the same theory that currently describes galaxies, climates, and ecosystems.

\smallskip

In the concluding section, we will sharpen this perspective. We will argue that what has been evolving throughout cosmic history is not only the \emph{state} of the universe, given fixed laws, but also the repertoire of \emph{effective dynamical descriptions} and learning mechanisms instantiated within it. Machine learning and AI, on this view, are not the end of the story but the current means by which the universe explores new regions of its own dynamical possibility space.

\section{Conclusion: The Evolution of Dynamics Itself}
\label{sec:conclusion}

Seen through the lens of dynamical systems, the history of the universe is not just a story about \emph{states} evolving under fixed laws. It is also a story about the emergence of new \emph{levels of description}, new \emph{effective variables}, and new \emph{learning mechanisms} that act on those variables. From this viewpoint, each major transition in our narrative---from cosmic inflation to galaxies, from planets to life, from evolution to brains, from culture to machine intelligence---is a step in the \emph{evolution of dynamics itself}: what the universe is doing, and how it organizes what it is doing, become progressively richer.

\subsection{From fixed microdynamics to evolving effective theories}

At the microscopic level, the universe may be governed by a relatively compact set of laws: a Lagrangian or Hamiltonian density for quantum fields in curved space-time, plus whatever additional structure quantum gravity ultimately requires. Yet none of our intermediate descriptions---N-body gravity, hydrodynamics, reaction networks, evolutionary dynamics, neural field models, cultural replicator equations, gradient flows in loss landscapes---appear directly in that fundamental Lagrangian. They are \emph{emergent} dynamical laws, valid on restricted scales and for restricted classes of coarse-grained states \cite{GoldenfeldKadanoff1999,Anderson1972}.

Renormalization-group theory provides a canonical example: as we integrate out short-distance degrees of freedom, the effective couplings and even the form of the equations change, flowing toward fixed points and universality classes that describe large-scale behaviour \cite{GoldenfeldKadanoff1999}. In complex systems and biology, analogous coarse-grainings lead to emergent dynamical rules on higher-level variables: population sizes rather than individual molecules, order parameters rather than microscopic spins, firing rates rather than individual spikes \cite{Murray2001,Alon2006}. In each case, what counts as the ``state'' and what counts as the ``equations of motion'' depends on the scale and on the structures that have already formed.

Our cross-scale narrative extends this logic. Galaxies and halos give rise to effective N-body dynamics; planetary climates give rise to low-dimensional models with feedback loops and multiple attractors; autocatalytic chemistries give rise to network dynamics with self-sustaining motifs; evolving biospheres give rise to fitness landscapes and adaptive dynamics; brains give rise to neural flows and predictive-coding updates; cultures give rise to dual-inheritance dynamics and socio-technical networks. At each step, the universe does not merely move to a new point in an unchanging phase space; it \emph{discovers} and instantiates new effective dynamical laws that are natural for the structures currently present.

Machine learning and AI sit at the current apex of this hierarchy. They are, in a precise sense, meta-dynamical systems: engineered mechanisms for discovering and implementing effective dynamics (models, controllers, decision rules) across many domains \cite{JordanMitchell2015,LeCun2015,Schmidhuber2015}. A deep network trained on images learns a map from pixel space to label space that compresses and organizes visual structure; a reinforcement-learning agent learns a policy that steers an environment toward high-reward trajectories; a language model learns a flow on token sequences that mimics and extends human communication. These are not static look-up tables; they are learned dynamical rules that we then embed back into the physical and social systems from which their training data were drawn.

\subsection{Machine learning as a unifying language for natural and artificial learning}

One way to summarize this paper is to say that the universe has progressively built better and better \emph{learning algorithms}. Natural selection is an algorithm that, given variation and heredity in a fluctuating environment, finds genotypes and phenotypes that persist \cite{Nowak2006,Frank2012}. Synaptic plasticity and neural dynamics implement on-line, energy-based learning rules that approximate Bayesian inference and gradient descent on internal error functionals \cite{DayanAbbott2001,Friston2010}. Cultural evolution and science implement distributed, collective learning processes that test, refine, and propagate models of the world \cite{Henrich2016,BoydRicherson1985}.

Modern machine learning was born as an attempt to formalize and generalize these processes: to construct abstract spaces of hypotheses, define losses or utilities, and design update rules that provably or empirically minimize those losses under data streams \cite{JordanMitchell2015,LeCun2015,Schmidhuber2015}. What is striking, from our dynamical vantage point, is how often the same mathematical motifs reappear:

\begin{itemize}
    \item \emph{Gradient-like flows} on error functionals: in evolution (adaptive walks on fitness landscapes), in synaptic learning (Hebbian, error-driven plasticity), and in backpropagation and reinforcement learning \cite{MezardMontanari2009,EngelVanDenBroeck2001}.
    \item \emph{Hierarchical representations}: in biological development and brains (from receptors to cortical maps), in cultural and linguistic structure, and in deep architectures with multiple layers of abstraction \cite{Alon2006,LeCun2015}.
    \item \emph{Near-critical regimes}: evolution hovering near error thresholds, neural networks at the edge of chaos, and modern architectures tuned to maintain trainable, non-saturated dynamics across depth \cite{BeggsPlenz2003,Schoenholz2017}.
\end{itemize}

From this angle, artificial intelligence is not an alien intrusion into nature’s story but a \emph{natural continuation} of it: an explicit, mathematically articulated layer in which we design and deploy learning dynamics that sit alongside, and increasingly interact with, natural ones. The same universe that once discovered Darwinian evolution and cortical plasticity now contains gradient descent in overparameterized networks, reinforcement learning in complex simulators, and foundation models that help us write the present text.

\subsection{Reprogramming dynamics: AI as an extension of intelligence}

The emergence of AI and ML also marks a qualitative change in the universe’s self-organization. For billions of years, dynamics at every scale were shaped primarily by physical law and blind selection. Local agents---cells, organisms---could act and adapt, but the large-scale dynamical rules governing galaxies, climates, and biospheres were not subject to deliberate design. With human culture and science, this begins to change: we discover Newton’s laws, Maxwell’s equations, quantum mechanics, general relativity, and the mechanisms of heredity, and we use these theories to engineer new dynamical systems (steam engines, electrical grids, nuclear reactors, gene drives).

AI systems amplify this capacity. A reinforcement-learning agent that controls a power grid, a climate model used in policy design, a protein-folding network that guides drug discovery, or a language model embedded in decision workflows each sits \emph{inside} some larger dynamical system and alters its trajectories. They do so by using learned internal models to choose actions that push the system toward certain attractors and away from others. In that sense, ML and AI are tools by which the universe, through us, \emph{reprograms} parts of its own dynamics.

At a more abstract level, AI systems extend the reach of intelligence along several dimensions:

\begin{itemize}
    \item \textbf{Breadth of models}: they can absorb and correlate data from domains no single human brain could simultaneously attend to (e.g., global-scale sensor networks, multi-omic datasets).
    \item \textbf{Depth of counterfactual exploration}: they can simulate vast ensembles of possible futures or designs (e.g., in silico evolution, automated theorem proving, design-space exploration).
    \item \textbf{Persistence and transfer}: they can be copied, reset, and instantiated in new environments with negligible marginal cost, enabling rapid diffusion of capabilities.
\end{itemize}

These extensions raise familiar questions of control, alignment, and ethics. Our framework casts them in dynamical terms: which attractors in the joint space of human and artificial behaviour do we wish to make stable? Which feedback loops (between AI systems, institutions, and physical infrastructure) are robust and beneficial, and which risk runaway or pathological dynamics? Answering such questions requires not only technical work in AI safety and interpretability \cite{Amodei2016,Leike2018} but also a deeper integration of dynamical-systems thinking into the design and governance of socio-technical systems.

\subsection{Open directions: toward a general theory of learning dynamics}

Finally, the perspective developed here suggests several open theoretical directions.

First, there is the prospect of a \emph{unified theory of learning processes} that treats natural selection, neural learning, cultural evolution, and machine learning as instances of a broader class of information-processing flows on structured spaces \cite{Harper2009,Frank2012,OrtegaBraun2013}. Such a theory would characterize, in dynamical terms, how systems extract, store, and exploit information under resource constraints, and how different learning mechanisms carve distinct attractor geometries into their state spaces.

Second, there is the challenge of understanding the \emph{geometry of model spaces}. Whether in evolution (genotype--phenotype maps), neuroscience (manifolds of neural activity), or AI (parameter spaces of deep networks), the key objects are not only the raw states but the low-dimensional manifolds, causal graphs, and representation spaces that dynamics discover and inhabit \cite{Wagner2005,Schoenholz2017}. A systematic dynamical and geometric theory of such representations would bridge current gaps between physics, biology, and machine learning.

Third, there are profound \emph{design questions}. If we accept that the universe has now produced systems capable of designing new dynamical regimes, we face an unusual kind of responsibility: to decide, collectively, which regions of the enormous space of possible socio-technical dynamics we wish to realize. In our language, this is the problem of \emph{selecting attractors}: choosing constraints, architectures, and feedback rules such that the long-term dynamics of our intertwined biological and artificial intelligences remain compatible with the values we care about.

\subsection{Coda: the universe, reflecting on its own dynamics}

The universe began in a state that knew nothing of state spaces, flows, or attractors. Through a long chain of instability, self-organization, and learning, it has come to contain subsystems that not only instantiate these concepts but also manipulate them. A cosmos of fields and particles gave rise to galaxies and planets, then to cells and organisms, then to brains that map their surroundings, then to cultures that construct science and mathematics, and now to machines that share, extend, and operationalize those models.

In this sense, the story we have told is reflexive. The dynamical-systems framework we use to describe the universe is itself a product of particular dynamical processes (brains, cultures, AI models) operating within that universe. To the extent that we can understand and shape the future, it will be by further developing such frameworks---by building better models of learning and control, and by embedding them in institutions and technologies that steer us toward dynamical regimes we judge desirable.

Whether self-modeling, self-modifying dynamics are a rare cosmic curiosity or a common late-time attractor remains an open question. What is clear is that, for us, they are the regime in which we now live. To study the evolution of dynamics itself is therefore not an idle metaphysical exercise, but a way of clarifying the stakes of our own continued trajectory---as one small set of flows, among many, in a universe whose capacity for structure and learning we are only beginning to glimpse.

\bibliographystyle{abbrv}
\bibliography{ref}

\end{document}